\documentclass[prd,preprint,showpacs,preprintnumbers,nofootinbib,eqsecnum,superscriptaddress]{revtex4}

 \usepackage[dvips,final]{graphicx}
  \usepackage{amssymb}
   \usepackage{amsmath}
    \usepackage{amsfonts}
     \usepackage{epsfig}
      \usepackage{bm}% bold math

\usepackage{mathpazo}

\usepackage[section]{placeins}

\usepackage{multirow}
\usepackage{ctable}
\usepackage{booktabs}
\usepackage{array}
\usepackage{tabularx}
\usepackage{xcolor}
\usepackage{pstricks}
\begin{document}

%\vfill
\title{Double-parton scattering effects in associated production\\ of charm mesons and dijets
at the LHC
}

\author{Rafa{\l} Maciu{\l}a}
\email{rafal.maciula@ifj.edu.pl} \affiliation{Institute of Nuclear
Physics, Polish Academy of Sciences, Radzikowskiego 152, PL-31-342 Krak{\'o}w, Poland}

\author{Antoni Szczurek\footnote{also at University of Rzesz\'ow, PL-35-959 Rzesz\'ow, Poland}}
\email{antoni.szczurek@ifj.edu.pl} \affiliation{Institute of Nuclear
Physics, Polish Academy of Sciences, Radzikowskiego 152, PL-31-342 Krak{\'o}w, Poland}

%\date{\today}

\begin{abstract}
We calculate several differential distributions for the production of charm and dijets.
Both single-parton scattering (SPS) and double-parton scattering (DPS) contributions are calculated
in the $k_T$-factorization approach. The Kimber-Martin-Ryskin unintegrated parton distributions
are used in our calculations. Relatively low cuts on jet transverse momenta are imposed to enhance 
the double-parton scattering mechanism contribution. We find dominance of the DPS contribution 
over the SPS one. We have found regions of 
the phase space where the SPS contribution 
is negligible compared to the DPS contribution.
The distribution in transverse momentum of charm quark/antiquark or
charmed mesons can be used to observe transition from the dominance of 
DPS at low transvsverse momenta to the dominance of SPS at large
transverse momenta. Very distinct azimuthal correlation patterns (for $c \bar{c}$, $c\mathrm{\textit{-jet}}$,
$\mathrm{\textit{jet-jet}}$, $D^0\mathrm{\textit{-jet}}$, $D^0 \overline{D^0}$) are predicted as a result of
the competition of the SPS and DPS mechanisms.

\end{abstract}


\maketitle

%-----------------------------------
\section{Introduction}
%-----------------------------------

The cross section for production of charm quarks or mesons is known to 
be very large especially at high energies which is caused by relatively 
small mass of the charm quark. On the other hand the mass of the charm
quark is large enough to use perturbative methods of quantum chromodynamics. 
Charm quarks are also produced abundantly in double \cite{Luszczak:2011zp,Maciula:2013kd,Cazaroto:2013fua,vanHameren:2014ava,Maciula:2016wci}
or multiple \cite{Maciula:2017meb,Cazaroto:2013fua} parton scattering.  The cross section for double 
$c \bar c$  production was shown to grow considerably with the collision
energy \cite{Luszczak:2011zp}. We have explained that total rates as well as several
differential distributions cannot be explained without inclusion of
double parton scattering. 
Many processes in association with charm quarks or mesons are possible
and can be studied at the LHC. Recently we discussed inclusive production
of single jet associated with $c \bar c$ or charmed mesons \cite{Maciula:2016kkx}.
Quite large cross sections were found there. Here we discuss inclusive
production of dijets in association with $c \bar c$ production. 

We wish to include both single parton scattering (SPS) and 
double parton scattering (DPS) processes.
We wish to discuss whether the process can be used to extract the
so-called $\sigma_{eff}$ parameter which governs the strength of double
parton scattering. The genral theoretical picture allows that this
quantity may depend on kinematical quantities as well as type of the process.
However, surprisingly similar values of $\sigma_{eff}$ were
obtained from different reactions. There are exceptions, much smaller
values were obtained for double $J/\psi$ charmonium production at 
large transverse momenta
\cite{Abazov:2014qba,Khachatryan:2014iia,Aaboud:2016fzt} but there the mechanism of the reaction is not yet fully 
understood \cite{Szczurek:2017uvc,CSSB2017}.
We focus on how to disantangle single and double parton scattering
contributions for a simultaneous production of $c \bar c$ (or charmed
mesons) and dijets.
In the current paper we present first predictions for the associated
production of charm and dijets as well as many differential distributions, 
many of them of the correlation character.

%-----------------------------------------------------------------
\section{A sketch of the theoretical formalism}
%-----------------------------------------------------------------

\subsection{Single-parton scattering}

%-----------------------------------------------------------------------------
\begin{figure}[!h]
\begin{minipage}{0.32\textwidth}
 \centerline{\includegraphics[width=1.0\textwidth]{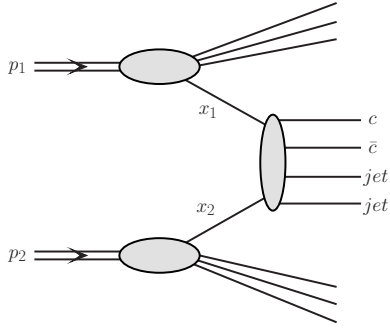}}
\end{minipage}
\caption{A diagrammatic representation of the SPS mechanism
            for the $pp \to c\bar c + \mathrm{2jets}\, X$ reaction.
\small 
 }
 \label{fig:diagrams_SPS}
\end{figure}
%------------------------------------------------------------------------------

Within the $k_T$-factorization approach the SPS cross section for 
$pp \to c\bar c + \mathrm{2jets}\, X$ reaction, sketched in Fig.~\ref{fig:diagrams_SPS}, can be written as
\begin{equation}
d \sigma_{p p \to c\bar c + \mathrm{2jets}} = \sum_{ij}
\int d x_1 \frac{d^2 k_{1t}}{\pi} d x_2 \frac{d^2 k_{2t}}{\pi}
{\cal F}_{i}(x_1,k_{1t}^2,\mu^2) {\cal F}_{j}(x_2,k_{2t}^2,\mu^2)
d {\hat \sigma}_{ij \to c \bar c + \mathrm{2part.} }
\; .
\label{cs_formula}
\end{equation}
In the formula above ${\cal F}_{i}(x,k_t^2,\mu^2)$ is a unintegrated
parton distribution function (uPDF) for a given type of parton $i = g, u, d, s, \bar u, \bar d, \bar s$. The uPDFs depend on longitudinal momentum fraction $x$, transverse momentum squared $k_t^2$ of the partons entering the hard process,
and in general also on a (factorization) scale of the hard process $\mu^2$.
The elementary cross section in Eq.~(\ref{cs_formula}) can be written
somewhat formally as:
\begin{equation}
d {\hat \sigma}_{ij \to c \bar c + \mathrm{2part.} } =
\prod_{l=1}^{4}
\frac{d^3 p_l}{(2 \pi)^3 2 E_l} 
(2 \pi)^4 \delta^{4}(\sum_{l=1}^{4} p_l - k_1 - k_2) \times\frac{1}{\mathrm{flux}} \overline{|{\cal M}_{i^* j^* \to c \bar c + \mathrm{2part.}}(k_{1},k_{2})|^2}
\; ,
\label{elementary_cs}
\end{equation}
where $E_{l}$ and $p_{l}$ are energies and momenta of final state particles. Above only dependence of the matrix element on four-vectors of incident partons $k_1$ and $k_2$ is made explicit. In general all four-momenta associated with partonic legs enter.
The matrix element takes into account that both partons entering the hard
process are off-shell with virtualities $k_1^2 = -k_{1t}^2$ and $k_2^2 = -k_{2t}^2$.
We take into account all 9 channels of the $2 \to 4$ type contributing to the cross section at the parton-level:\\
\begin{center}
$\#1 = g \; g \to g \; g \; c \; \bar{c}$ $\;\;\;\;\;\;$ $\#2 = g \; g \to q \; \bar{q} \; c \; \bar{c}$
$\;\;\;\;\;\;$ $\#3 = g \; q \to g \; q \; c \; \bar{c}$\\
$\#4 = q \; g \to q \; g \; c \; \bar{c}$ $\;\;\;\;\;\;$ $\#5 = q \; \bar{q} \to q' \; \bar{q}' \; c \; \bar{c}$
$\;\;\;\;\;\;$ $\#6 = q \; \bar{q} \to g \; g \; c \; \bar{c}$\\
$\#7 = q \; q \to q \; q \; c \; \bar{c}$ $\;\;\;\;\;\;$ $\#8 = q \; q' \to q \; q' \; c \; \bar{c}$
$\;\;\;\;\;\;$ $\#9 = q \; \bar{q} \to q \; \bar{q} \; c \; \bar{c}$.
\end{center}

The off-shell matrix elements are well known only in the leading-order (LO) and only for limited types of QCD $2 \to 2$ processes (see \textit{e.g.} heavy quarks \cite{Catani:1990eg}, dijet \cite{Nefedov:2013ywa}, Drell-Yan \cite{Nefedov:2012cq}). Some first steps to calculate NLO corrections in the $k_{T}$-factorization framework have been done only very recently for diphoton production \cite{Nefedov:2015ara,Nefedov:2016clr}.
For higher final state parton multiplicities, relevant amplitudes can be calculated analytically applying suitably defined Feynman rules \cite{vanHameren:2012if} or recursive methods, like  generalised  BCFW recursion \cite{vanHameren:2014iua}, or numerically with the help of methods of numerical BCFW recursion \cite{Bury:2015dla}. The latter method was already successfully applied for $2 \to 3$ production mechanisms in the case of $c\bar c + \mathrm{jet}$ \cite{Maciula:2016kkx} and even for $2 \to 4$ processes in the case of $c\bar c c\bar c$ \cite{vanHameren:2015wva} and four-jet \cite{Kutak:2016mik} final states.

In this paper we follow the same numerical techniques. The calculation has been performed with the help of KaTie \cite{vanHameren:2016kkz}, which is a complete Monte Carlo parton-level event generator for hadron scattering processes. It can can be applied to any arbitrary processes within the Standard Model, for up to four final-state particles and beyond, and for any initial-state partons on-shell or off-shell. The scattering amplitudes are calculated numerically as a function of the external four-momenta via Dyson-Schwinger recursion \cite{Caravaglios:1995cd} generalized also to tree-level off-shell amplitudes. The phase space integration in KaTie is done with the help of a full Monte Carlo program with an adaptive phase space generator, previously incorporated as a part of the AVHLIB library ~\cite{vanHameren:2007pt,vanHameren:2010gg}, that deals with the integration variables related to both the initial-state momenta and the final-state momenta. KaTie can be used for single-parton scattering as well as
for multi-parton scattering processes. 

In the present calculation, we use $\mu^2 \! = \! \frac{m_{1t}^{2}+m_{2t}^{2}+p_{3t}^{2}+p_{4t}^{2}}{4}$ as the renormalization/factorization scale, where $m_{1t}, m_{2t}$ are the transverse mass of the outgoing $c$-quark and $\bar c$-antiquark and $p_{3t}, p_{4t}$ are the transverse momenta of outgonig jets. Furthermore, we take running $\alpha_{s}$ at next-to-leading order (NLO) and
charm quark mass $m_c$ = 1.5 GeV. The parameters are the same for both $k_t$-factorization and in the reference
collinear-factorization calculations. Uncertainties related to the choice of the parameters were discussed
\textit{e.g.} in Ref.~\cite{vanHameren:2014ava} and will be not considered here. We use the Kimber-Martin-Ryskin (KMR) \cite{Kimber:2001sc,Watt:2003vf} unintegrated distributions for quarks and gluon calculated from the MMHT2014nlo PDFs \cite{Harland-Lang:2014zoa}. The above choices are kept the same also in the case of double-parton scattering calculation except of the scales.

\subsection{Double-parton scattering}

According to the general form of the multiple-parton scattering theory (see \textit{e.g.} Refs.~\cite{Diehl:2011tt,Diehl:2011yj})
the DPS cross sections can be expressed in terms of the double parton distribution functions (dPDFs).
These objects should fulfill sum rules and take into account all the correlations between the two partons, including transverse and longitudinal momenta correlations as well as color, flavour and spin correlations. The theory of dPDFs is well established but still not fully applicable for phenomenological studies. The currently available models of the dPDFs are still rather 
at a preliminary stage. So far they are formulated exlusively for gluon or for valence quarks and only in a leading order framework
which may be not sufficient for many processes, especially when charm production is considered.    

Instead of the general form, one usually follows the assumption of the factorization of the DPS cross section.
Within this framework, the dPDFs are taken in the following factorized form:  
\begin{equation}
D_{1, 2}(x_1,x_2,\mu) = f_1(x_1,\mu)\, f_2(x_2,\mu) \, \theta(1-x_1-x_2) \, ,
\end{equation}
where $D_{1, 2}(x_1,x_2,\mu)$ is the dPDF and
$f_i(x_i,\mu)$ are the standard single PDFs for the two generic partons in the same proton. The factor $\theta(1-x_1-x_2)$ ensures that
the sum of the two parton momenta does not exceed 1. 
%
%-----------------------------------------------------------------------------
\begin{figure}[!h]
\begin{minipage}{0.32\textwidth}
 \centerline{\includegraphics[width=1.0\textwidth]{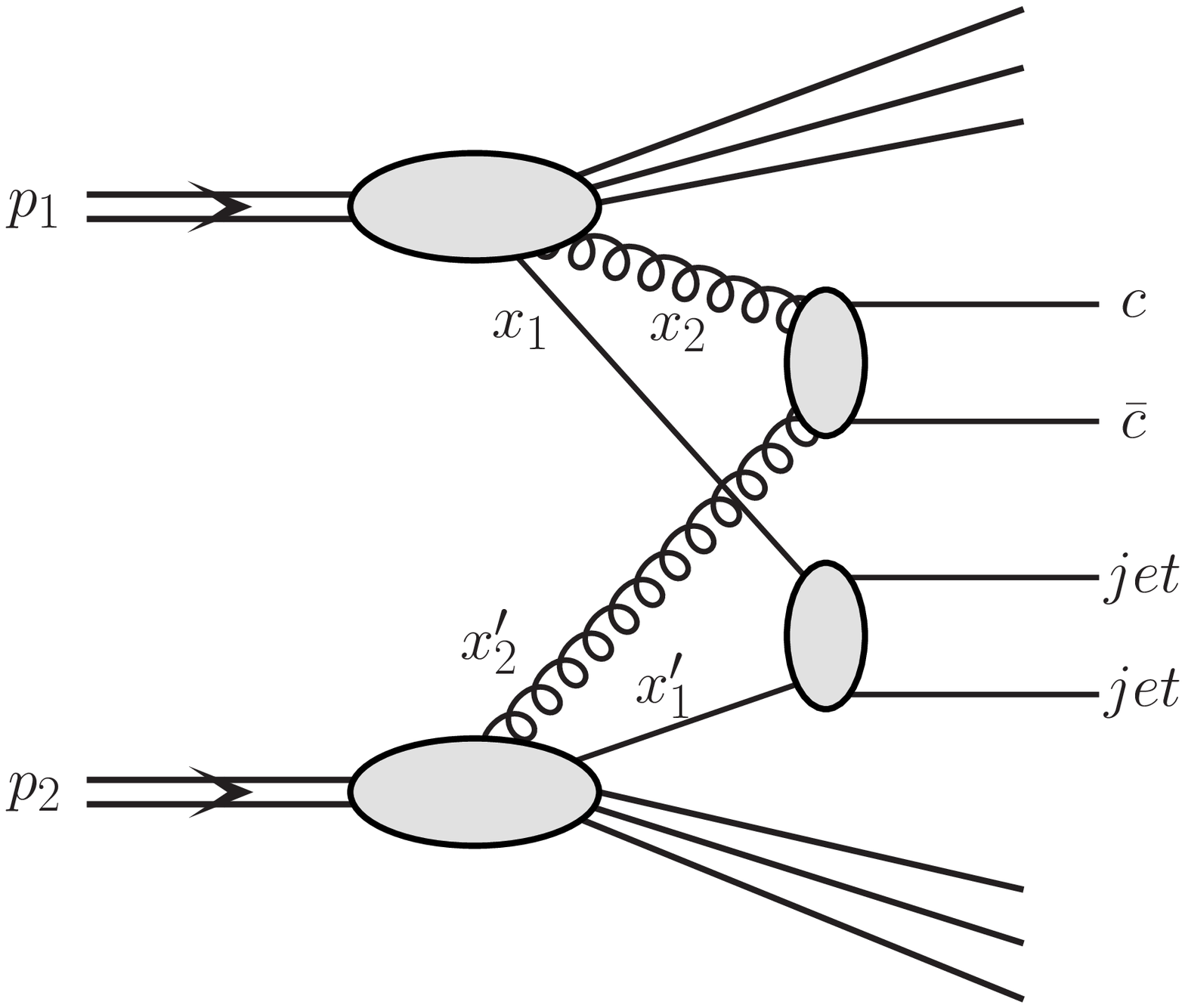}}
\end{minipage}
\hspace{0.1cm}
\begin{minipage}{0.32\textwidth}
 \centerline{\includegraphics[width=1.0\textwidth]{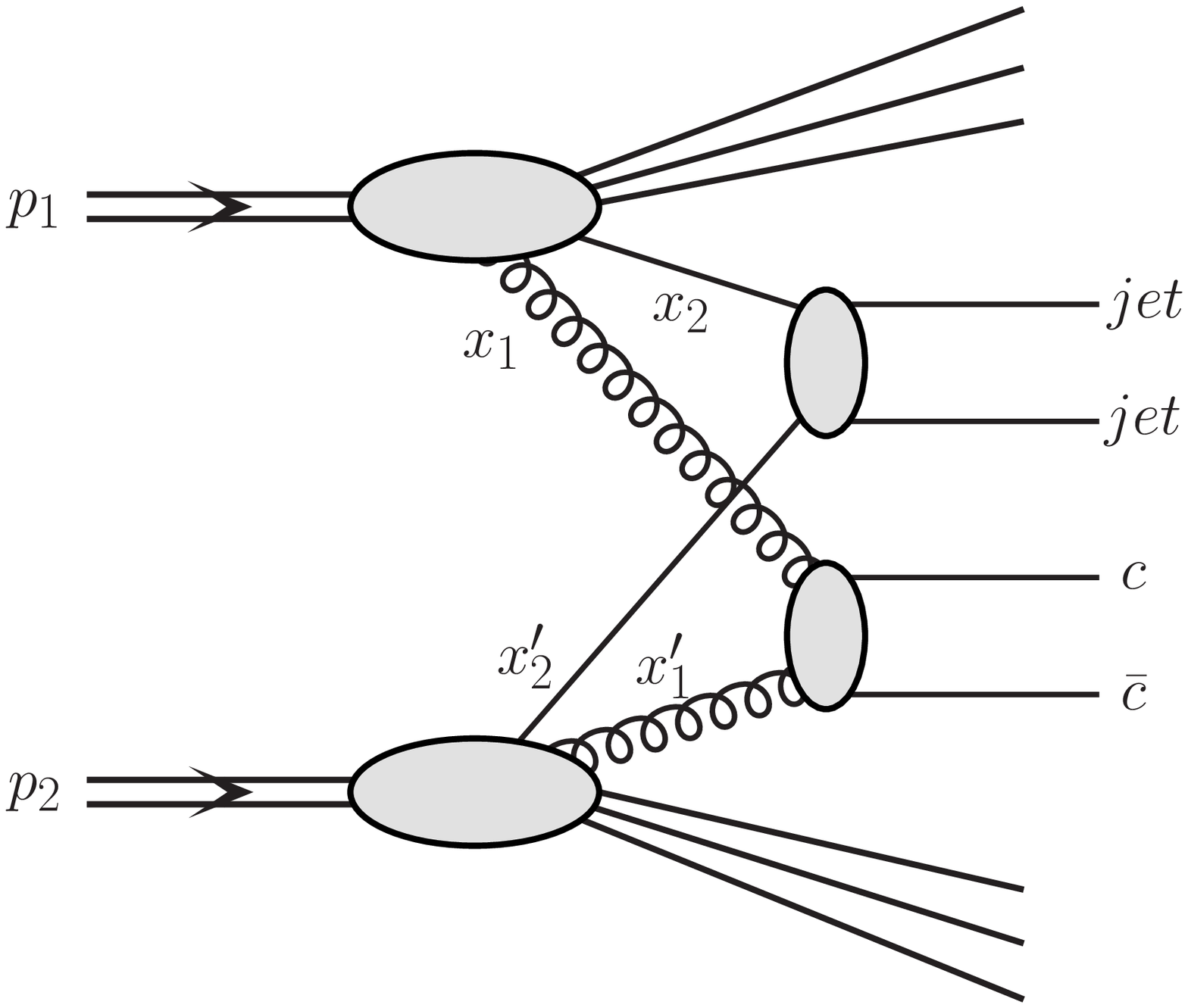}}
\end{minipage}
   \caption{A diagrammatic representation of the DPS mechanisms
            for the $pp \to c\bar c + \mathrm{2jets}\, X$ reaction.
\small 
 }
 \label{fig:diagrams_DPS}
\end{figure}
%------------------------------------------------------------------------------

According to the above, the differential cross section for $pp \to c\bar c + \mathrm{2jets}\; X$ reaction within the DPS mechanism, sketched in Fig.~\ref{fig:diagrams_DPS}, can be expressed as follows: 
\begin{equation}
\frac{d\sigma^{DPS}(c \bar c + \mathrm{2jets})}{d\xi_{1}d\xi_{2}} = \sum_{i,j} \;  \frac{1}{\sigma_{eff}} \cdot \frac{d\sigma^{SPS}(g g \to c \bar c)}{d\xi_{1}} \! \cdot \! \frac{\sigma^{SPS}(i j \to \mathrm{2jets})}{d\xi_{2}},
\label{basic_formula}
\end{equation}
where $\xi_{1}$ and $\xi_{2}$ stand for generic phase space kinematical variables for the first and second scattering, respectively.
When integrating over kinematical variables one recovers the commonly used pocket-formula:
\begin{equation}
\sigma^{DPS}(c \bar c + \mathrm{2jets}) = \sum_{i,j} \;  
\frac{\sigma^{SPS}(g g \to c \bar c) \! \cdot \! \sigma^{SPS}(i j \to \mathrm{2jets})}{\sigma_{eff}}\; .
\label{basic_formula}
\end{equation}

The effective cross section $\sigma_{eff}$ provides a proper normalization of the DPS cross section and can be roughly interpreted 
as a measure of the transverse correlation of the two partons inside 
the hadrons. The longitudinal parton-parton correlations should become far less
important as the energy of the collision is increased, due to the increase in the parton multiplicity. It is belived that for small-$x$ partons and for low and intermediate scales the possible longitudinal correlations can be safely
neglected (see \textit{e.g.} Ref.~\cite{Gaunt:2009re}). 
In this paper we use world-average value of $\sigma_{eff} = 15$ mb provided by 
several experiments at Tevatron \cite{Abe:1997bp,Abe:1997xk,Abazov:2009gc} and LHC \cite{Aaij:2012dz,Aad:2013bjm,Chatrchyan:2013xxa,Aad:2014rua,Aaboud:2016dea}. Future experiments may verify this value.

There are several effects that may lead to a violation of the factorized anstaz, which is a severe approximation.
The flavour, spin and color correlations lead, in principle, to interference effects
that result in breaking of the pocket-formula (see \textit{e.g.} Refs.~\cite{Diehl:2011tt,Diehl:2011yj}. In any case, the spin polarization of the two partons from one hadron
can be mutually correlated, especially when the partons are relatively close in phase space (having comparable $x$'s). The two-parton distributions have a nontrivial color structure which also may lead to a non-negligible correlations effects. 
Such effects are usually not included in phenomenological analyses. They were exceptionally discussed in the context of double charm production \cite{Echevarria:2015ufa} but in this case the corresponding effects were found to be very small.
Moreover, including perturbative parton splitting mechanism \cite{Ryskin:2011kk,Gaunt:2012dd,Gaunt:2014rua} and/or imposing sum rules \cite{Golec-Biernat:2015aza} also leads to a breaking of the pocket-formula.  
However, taken the above and looking forward to further improvements in this field, we choose to
limit ourselves to a more pragmatic approach in this paper.

In our present analysis cross sections for each step of the DPS mechanism are calculated in the
$k_T$-factorization approach, that is:
\begin{eqnarray}
\frac{d \sigma^{SPS}(p p \to c \bar c \; X_1)}{d y_1 d y_2 d^2 p_{1,t} d^2 p_{2,t}} 
&& = \frac{1}{16 \pi^2 {\hat s}^2} \int \frac{d^2 k_{1t}}{\pi} \frac{d^2 k_{2t}}{\pi} \overline{|{\cal M}_{g^{*} g^{*} \rightarrow c \bar{c}}|^2} \nonumber \\
&& \times \;\; \delta^2 \left( \vec{k}_{1t} + \vec{k}_{2t} - \vec{p}_{1t} - \vec{p}_{2t}
\right)
{\cal F}_{g}(x_1,k_{1t}^2,\mu^2) {\cal F}_{g}(x_2,k_{2t}^2,\mu^2),
\nonumber
\end{eqnarray}
\begin{eqnarray}
\frac{d \sigma^{SPS}(p p \to \mathrm{2jets} \; X_2)}{d y_3 d y_4 d^2 p_{3,t} d^2 p_{4,t}} 
&& = \frac{1}{16 \pi^2 {\hat s}^2} \sum_{ij} \int \frac{d^2 k_{3t}}{\pi} \frac{d^2 k_{4t}}{\pi} \overline{|{\cal M}_{i^{*} j^{*} \rightarrow \mathrm{2part.}}|^2} \nonumber \\
&&\times \;\; \delta^2 \left( \vec{k}_{3t} + \vec{k}_{4t} - \vec{p}_{3t} - \vec{p}_{4t}
\right)
{\cal F}_{i}(x_3,k_{3t}^2,\mu^2) {\cal F}_{j}(x_4,k_{4t}^2,\mu^2). \nonumber \\
\end{eqnarray}
The numerical calculations for both SPS mechanisms are also done within the KaTie code, where the relevant fully gauge-invariant off-shell $2 \to 2$ matrix elements ${\cal M}_{g^{*} g^{*} \rightarrow c \bar{c}}$ and ${\cal M}_{i^{*} j^{*} \rightarrow \mathrm{2part.}}$ are obtained numerically. Their useful analytical form can be found in Ref.~\cite{Catani:1990eg} for $c \bar c$ and in Ref.~\cite{Nefedov:2013ywa} for dijet production.  
Here, the strong coupling constant $\alpha_S$ and uPDFs are taken the same as in the case of the calculation of the SPS mechanism for $c\bar c + \mathrm{2jets}$ production. The factorization and renormalization scales for the two single scatterings are $\mu^2 \! = \! \frac{m_{1t}^{2}+m_{2t}^{2}}{2}$ for the first, and $\mu^2 \! = \! \frac{p_{3t}^{2}+p_{4t}^{2}}{4}$ for the second subprocess.
The framework of the $k_T$-factorization approach together with the KMR uPDFs was used with success in describing inclusive spectra of
$D$, $D\bar D$ correlations \cite{Maciula:2013wg,Karpishkov:2016hnx} as well as in the case of dijet production \cite{Nefedov:2013ywa} and therefore can be expected to be a good starting point for the DPS predictions for the $c\bar c + \mathrm{2jets}$ final state.

%--------------------------------------------------
\section{Numerical results}
%--------------------------------------------------

\subsection{$\bm{c \bar c + \mathrm{2jets}}$}

We start presentation of the numerical predictions with the results for production of charm quark-antiquark pair associated with two jets at $\sqrt{s} = 13$ TeV.
Here, the phase space for charm quarks is almost unlimited, with broad range of rapidities $|y_{c}| < 8$ and without any cuts on their transverse momenta. For jets we keep the kinematical regime relevant for the ATLAS/CMS experiments, with $|y_{\mathrm{jet}}| < 4.9$ and with transverse momentum cut
$p_{T}^{\mathrm{jet}} > p_{T,\mathrm{cut}}^{\mathrm{jet}} = 20$ GeV for leading and subleading jet.    

%-----------------------------------------------------------------------------
\begin{figure}[!h]
\begin{minipage}{0.47\textwidth}
 \centerline{\includegraphics[width=1.0\textwidth]{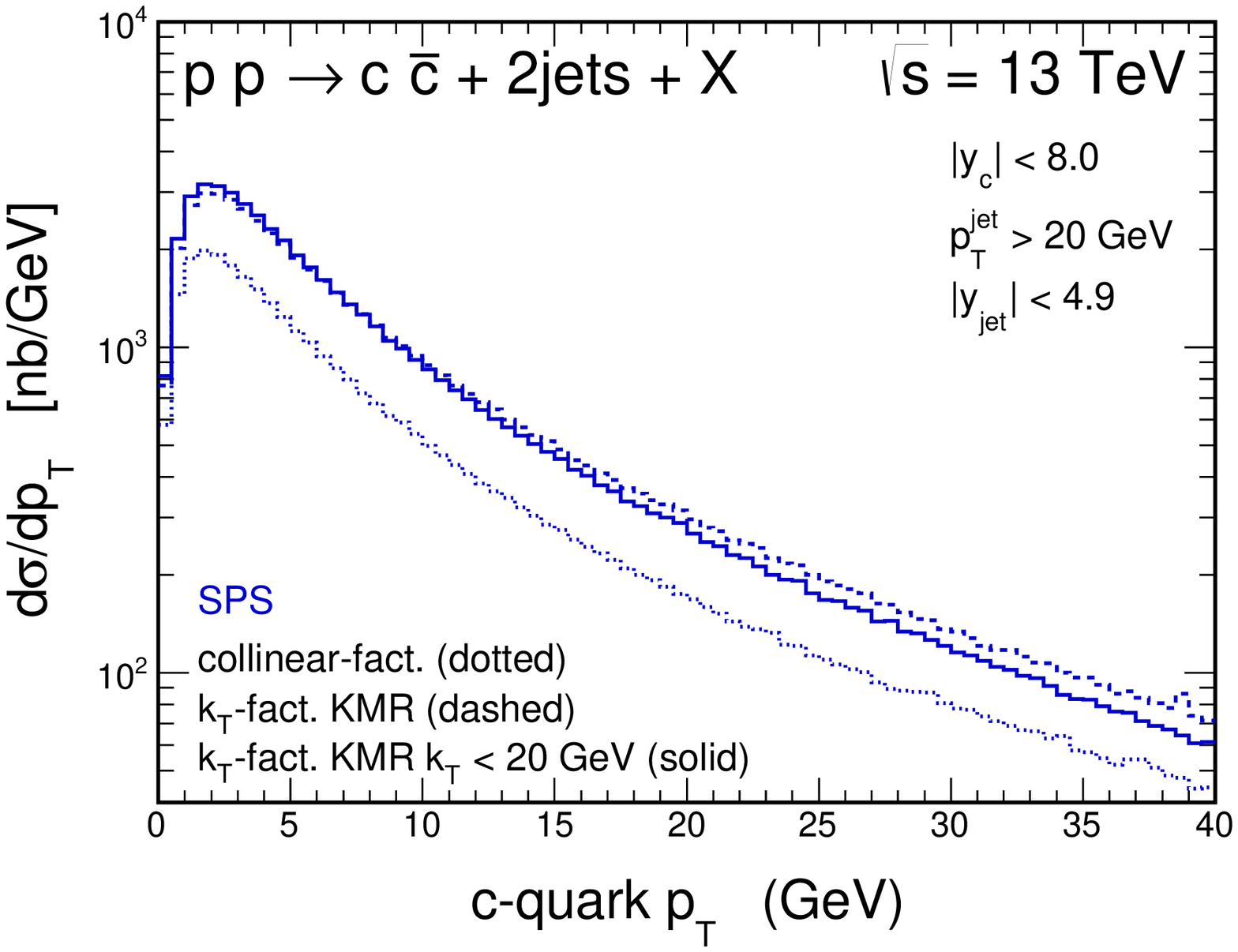}}
\end{minipage}
\hspace{0.5cm}
\begin{minipage}{0.47\textwidth}
 \centerline{\includegraphics[width=1.0\textwidth]{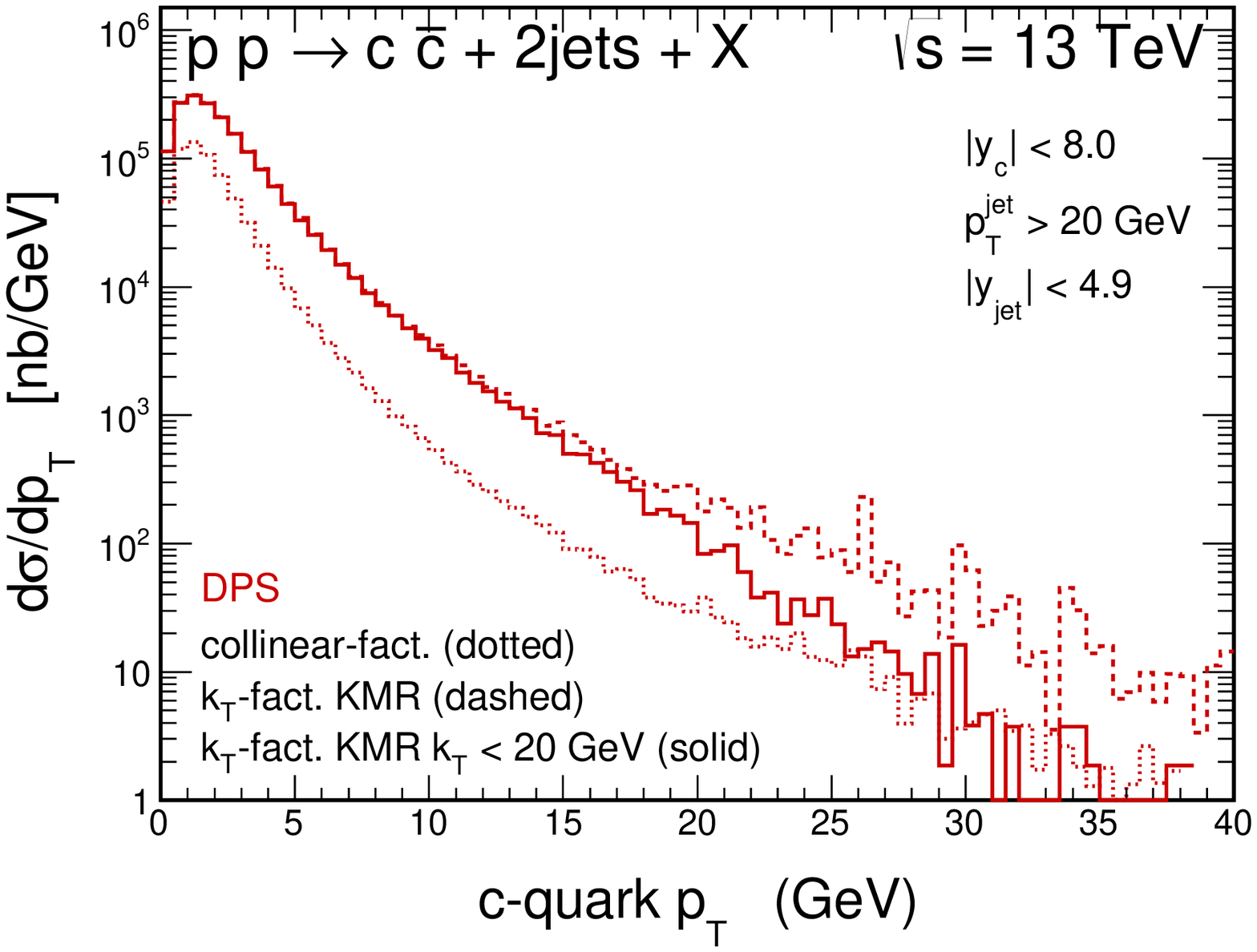}}
\end{minipage}
   \caption{
\small The transverse momentum distribution of charm quarks/antiquarks for SPS (left) and DPS (right) mechanisms at $\sqrt{s} = 13$ TeV.
The three different histograms correspond to different approaches used in the calculations. Details are specified in the figure.}
 \label{fig:pt_c_compar}
\end{figure}
%------------------------------------------------------------------------------

In Fig.~\ref{fig:pt_c_compar} we show transverse momentum distributions of charm quark/antiquark for single-parton scattering (left panel) and for double-parton scattering (right panel) mechanism at $\sqrt{s} = 13$ TeV. The three different histograms correspond to different approaches used in the calculations: LO collinear approximation (dotted), $k_{T}$-factorization (dashed) and $k_{T}$-factorization with extra cut on incident parton transverse momenta $k_{T} < 20$ GeV (solid). The KMR model for uPDFs, due to its special construction, allows for additional emission of hard gluon or quark from uPDF when the initial parton that enters the hard scattering has transverse momentum $k_{T} > \mu$. To make predictions for the final state with $c \bar c$-pair and with exactly two jets one needs to introduce the special limitation: $k_{T} < p_{T,\mathrm{cut}}^{\mathrm{jet}}$. We see that the effects related with this cut become important only when going to larger transverse momenta of charm quark and are much stronger in the case of the DPS results, where the cut is applied for all four incident partons. In both cases, for the SPS and the DPS calculation, the LO collinear approximation leads to a significantly smaller cross sections than those obtained within the $k_T$-factorization in the whole considered range of charm quark transverse momenta. 

%-----------------------------------------------------------------------------
\begin{figure}[!h]
\begin{minipage}{0.47\textwidth}
 \centerline{\includegraphics[width=1.0\textwidth]{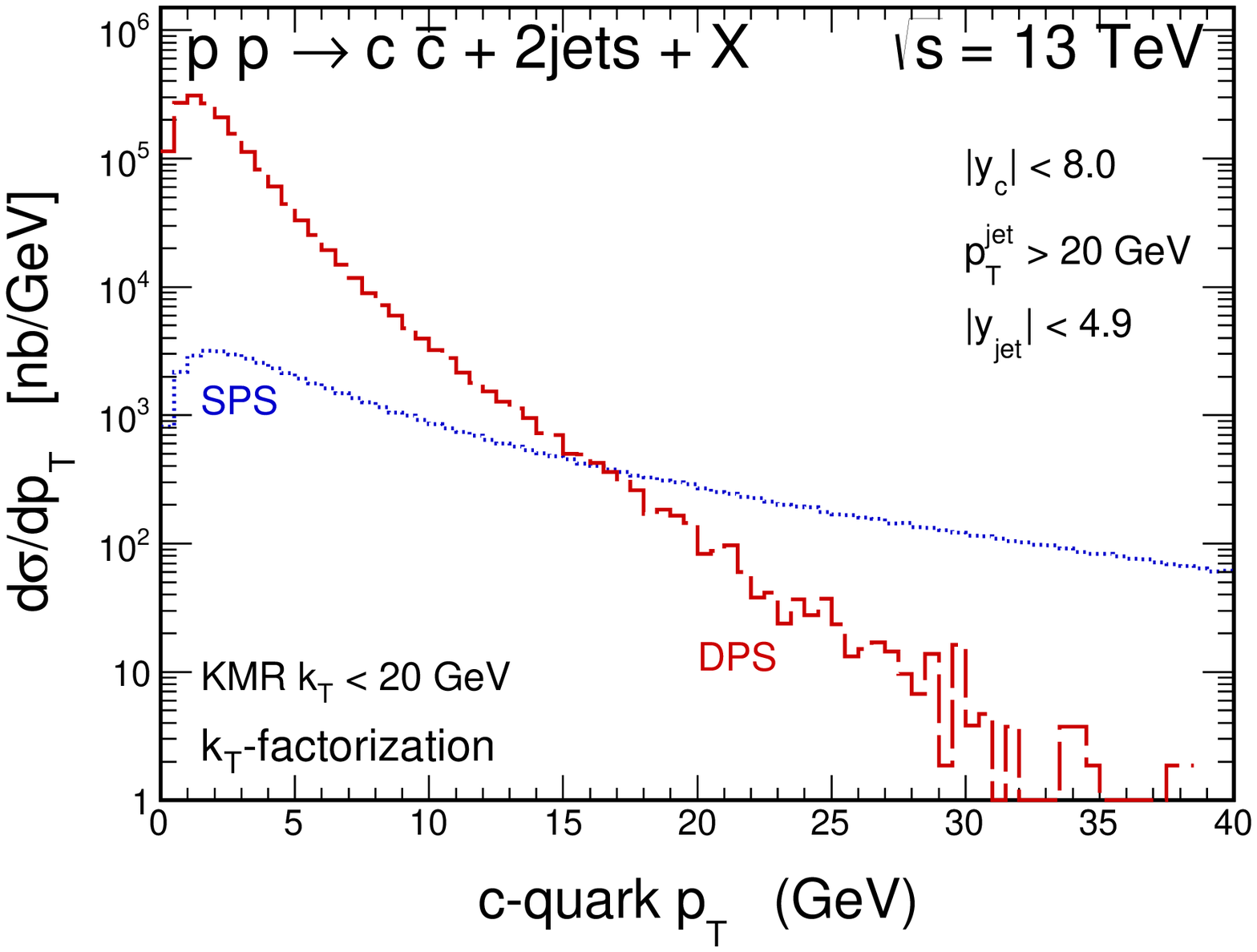}}
\end{minipage}
\hspace{0.5cm}
\begin{minipage}{0.47\textwidth}
 \centerline{\includegraphics[width=1.0\textwidth]{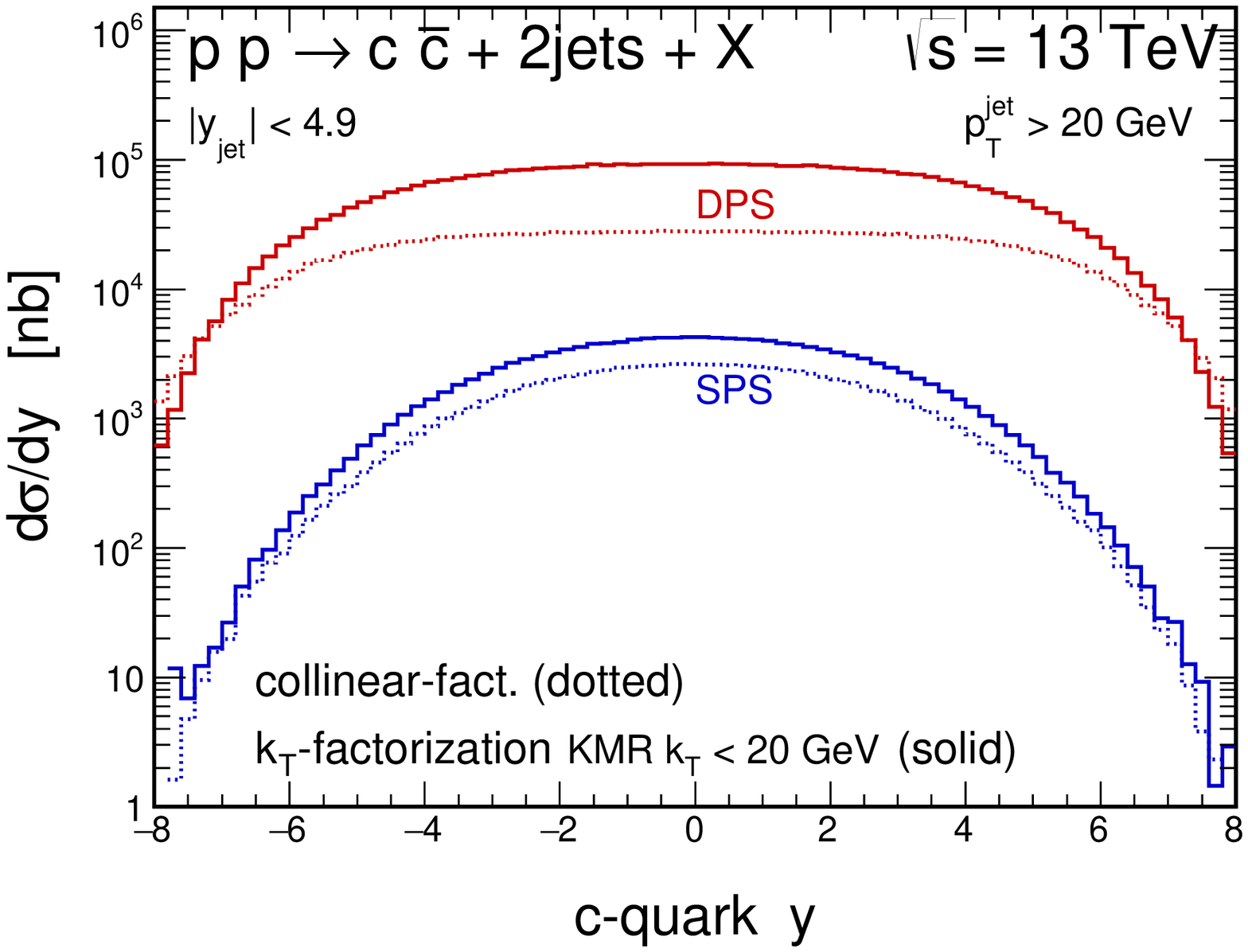}}
\end{minipage}
   \caption{
\small Left panel: The transverse momentum distribution of charm quarks/antiquarks for SPS (dotted) and DPS (dashed) mechanisms calculated within the $k_{T}$-factorization approach for the KMR uPDFs and with the $k_{T} < p_{T,\mathrm{cut}}^{jet}$ constrain. Right panel: The rapidity distribution of charm quarks/antiquarks for SPS (lower) and DPS (upper)
mechanisms calculated within the collinear (dotted) and the $k_{T}$-factorization (solid) approaches. Details are specified in the figure.
 }
 \label{fig:pt_y_c}
\end{figure}
%------------------------------------------------------------------------------

In the left panel of Fig.~\ref{fig:pt_y_c} we show again the transverse momentum distribution of charm quark/antiquark. Here, the DPS (dashed histogram) and the SPS (dotted histogram) contributions calculated within the $k_T$-factorization approach are shown together on the same plot.
The DPS contribution clearly dominates over the SPS one in the region of $c$-quark $p_{T} < 15$ GeV. In the right panel of Fig.~\ref{fig:pt_y_c}
we present rapidity distribution of charm quark/antiquark. The two upper histograms correspond to the DPS mechanism and the two lower histograms   
correspond to the SPS contribution. Here, results for the $k_{T}$-factorization approach (solid histograms) are shown together with results obtained with the LO collinear approach (dotted histograms). The DPS component significantly dominates in the whole considered range of rapidities and the relative contribution of the SPS mechanism becomes even smaller when moving to forward/backward region.

\begin{figure}[!h]
\begin{minipage}{0.47\textwidth}
 \centerline{\includegraphics[width=1.0\textwidth]{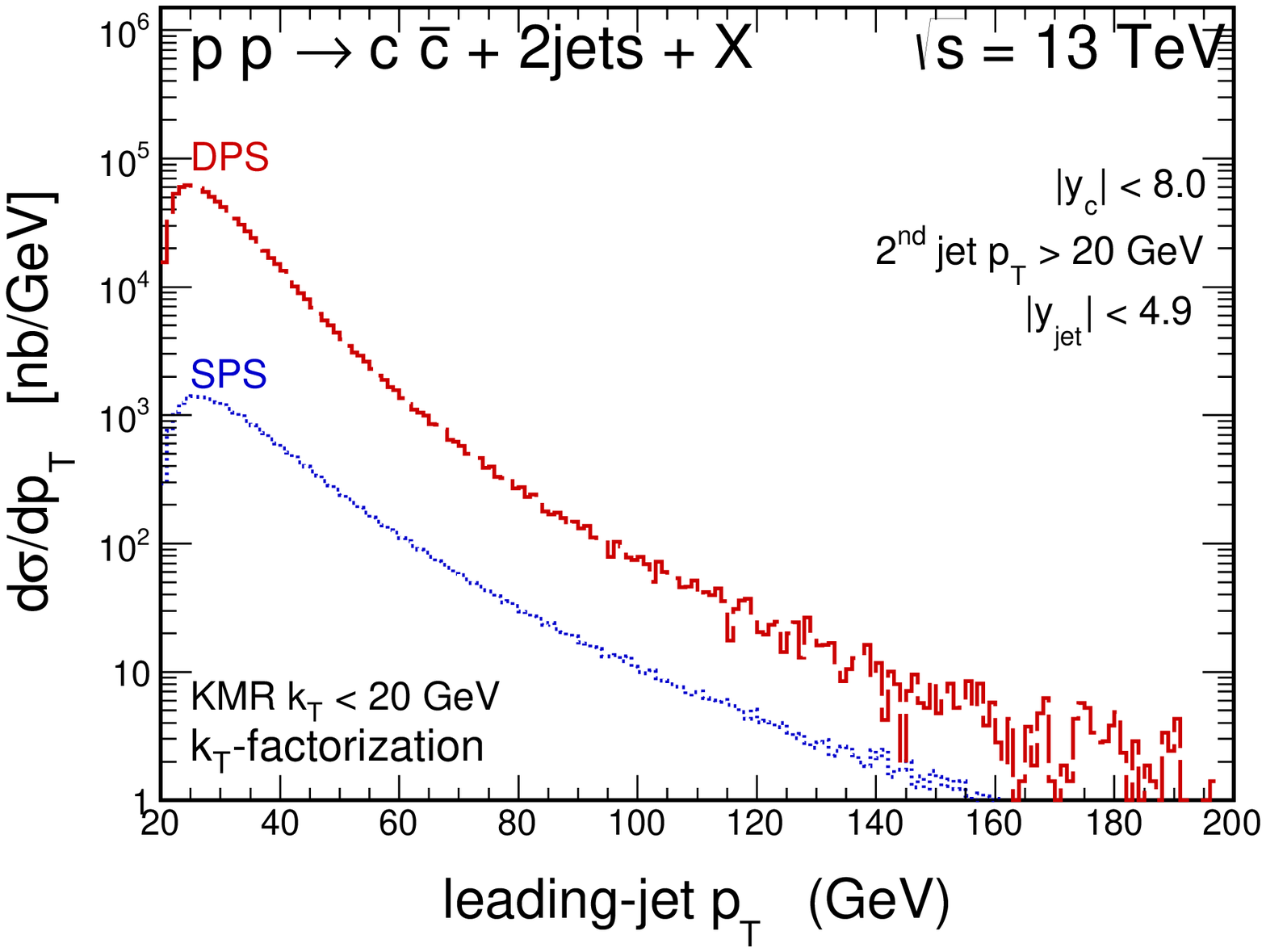}}
\end{minipage}
\hspace{0.5cm}
\begin{minipage}{0.47\textwidth}
 \centerline{\includegraphics[width=1.0\textwidth]{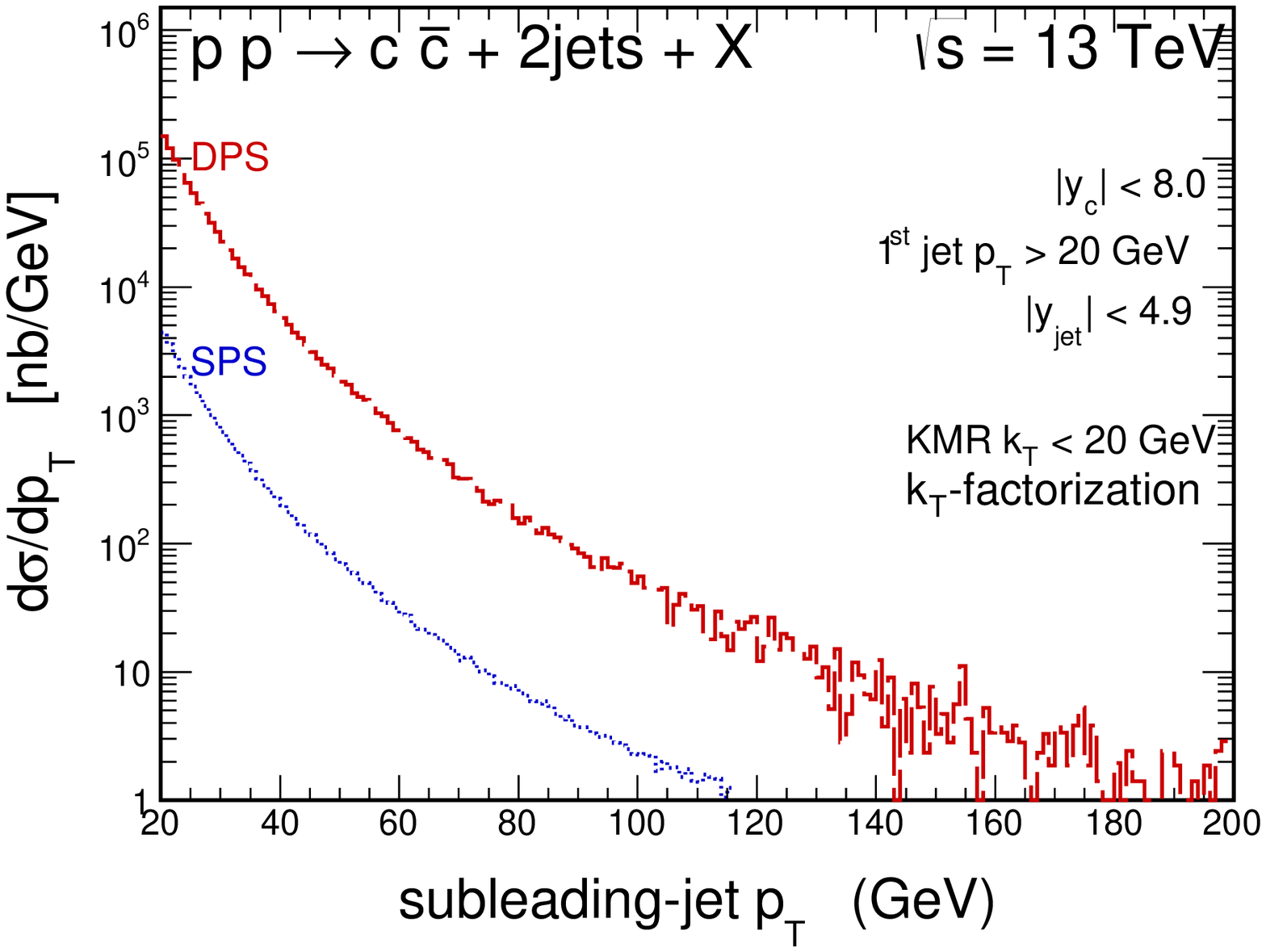}}
\end{minipage}
   \caption{
\small The transverse momentum distribution of the leading (left) and the subleading (right) jet for SPS (dotted) and DPS (dashed) mechanisms calculated within the $k_{T}$-factorization approach for the KMR uPDFs and with the $k_{T} < p_{T,\mathrm{cut}}^{jet}$ constrain. Details are specified in the figure.
 }
 \label{fig:pt_jets}
\end{figure}

Figure~\ref{fig:pt_jets} shows transverse momentum distributions of the leading (left panel) and the subleading (right panel) jet calculated in the $k_{T}$-factorization approach for the SPS (lower histograms) and for the DPS mechanism (upper histograms). The DPS component dominates in the whole range of the considered jet transverse momenta.

%-----------------------------------------------------------------------------
%------------------------------------------------------------------------------

\begin{figure}[!h]
\begin{minipage}{0.47\textwidth}
 \centerline{\includegraphics[width=1.0\textwidth]{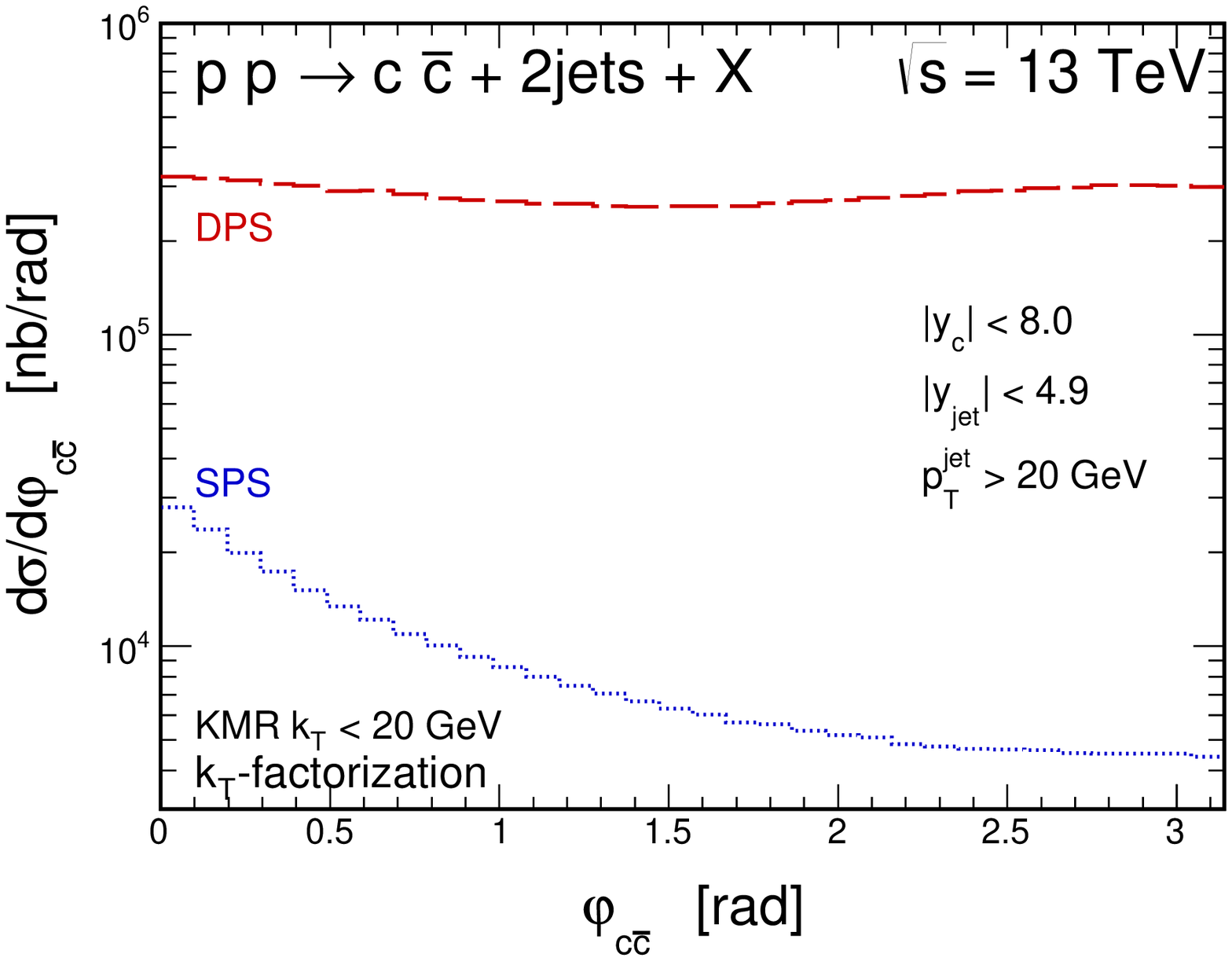}}
\end{minipage}
\hspace{0.5cm}
\begin{minipage}{0.47\textwidth}
 \centerline{\includegraphics[width=1.0\textwidth]{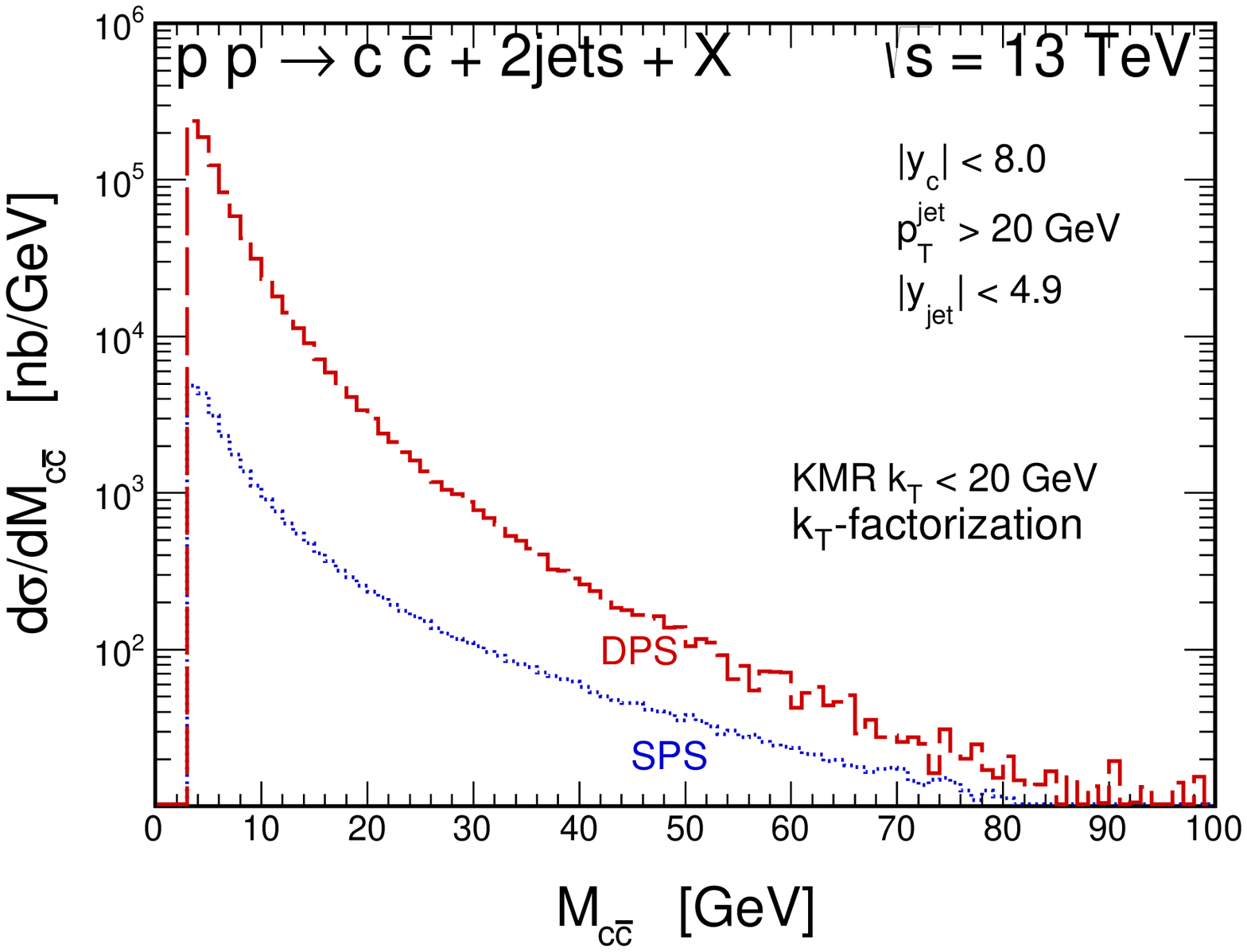}}
\end{minipage}
   \caption{
\small The same as in Fig.~\ref{fig:pt_jets} but for the azimuthal angle $\varphi_{c\bar{c}}$ between the $c$-quark and the $\bar{c}$-antiquark (left)
and for the $c\bar{c}$-pair invariant mass $M_{c\bar{c}}$ (right).
 }
 \label{fig:corr_ccbar}
\end{figure}

In Fig.~\ref{fig:corr_ccbar} we present some $c\bar c$ correlation observables. The left panel shows distribution in azimuthal angle $\varphi_{c\bar{c}}$ between the $c$-quark and the $\bar{c}$-antiquark. Again, the DPS mechanism dominates over the SPS one in the whole range of relative azimuthal angle. The shape of the distribution for the DPS is determined directly by the inclusive single $c\bar c$-pair production mechanism.
In the case of the SPS component we observe an evident enhancement of the cross section in the region of rather small angles, which is not a typical behaviour \textit{e.g.} in the case of inclusive charm production. In the right panel of Fig.~\ref{fig:corr_ccbar} we present differential cross section as a function of invariant mass of the $c\bar c$-system $M_{c\bar c}$. Both mechanisms have different slope of the distribution. At small invariant masses the DPS component clearly dominates over the SPS one. In the region of $M_{c\bar c} < 20$ GeV the difference is bigger than one order of magnitude. Both of the mechanisms become comparable starting from $M_{c\bar c} \approx 70$ GeV.     
   
%-----------------------------------------------------------------------------
%------------------------------------------------------------------------------

\begin{figure}[!h]
\begin{minipage}{0.47\textwidth}
 \centerline{\includegraphics[width=1.0\textwidth]{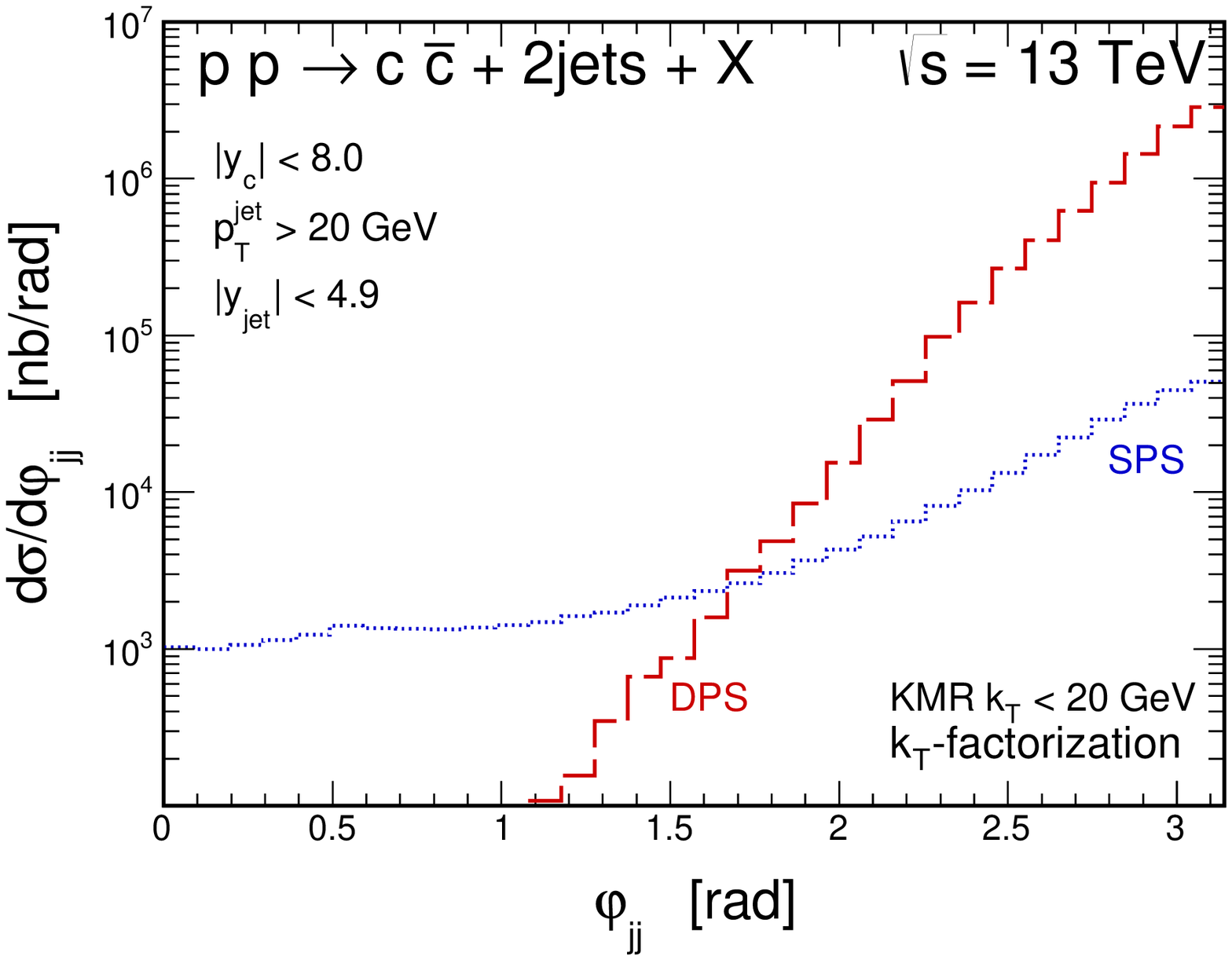}}
\end{minipage}
\hspace{0.5cm}
\begin{minipage}{0.47\textwidth}
 \centerline{\includegraphics[width=1.0\textwidth]{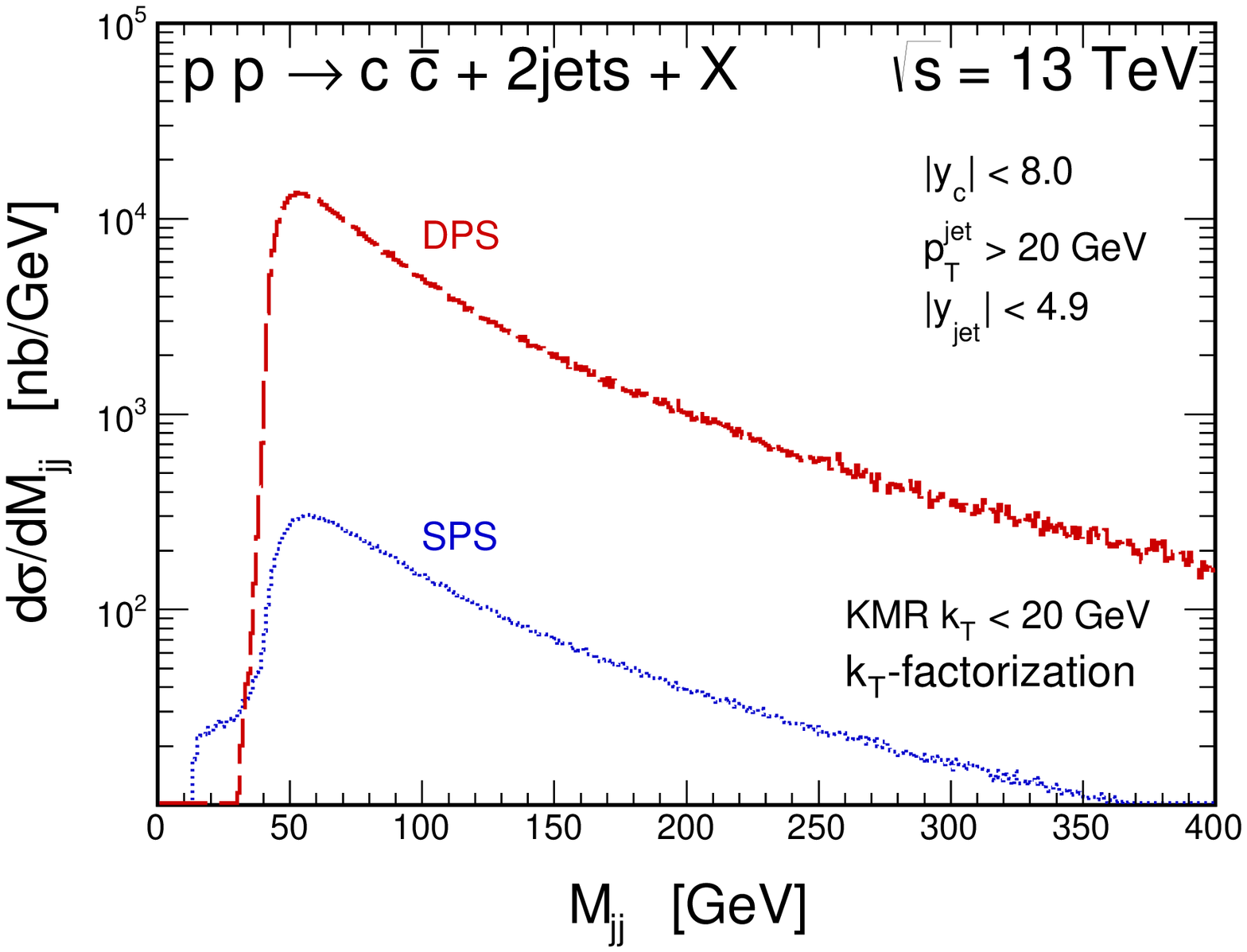}}
\end{minipage}
   \caption{
\small The same as in Fig.~\ref{fig:pt_jets} but for the azimuthal angle $\varphi_{jj}$ between the two jets (left)
and for the dijet invariant mass $M_{jj}$ (right).
 }
 \label{fig:corr_jets}
\end{figure}

Figure~\ref{fig:corr_jets} presents differential distributions as a function of the azimuthal angle $\varphi_{jj}$ between the two jets (left panel)
and as a function of the dijet invariant mass $M_{jj}$ (right panel). The SPS component is much more decorrelated in azimuthal angle than the DPS one.
Both of them have a similar shape of the dijet invariant mass and differ only as far as only the normalization is considered. The DPS mechanism dominates in the whole range of the dijet invariant mass. 

%-----------------------------------------------------------------------------
%------------------------------------------------------------------------------

\begin{figure}[!h]
\begin{minipage}{0.47\textwidth}
 \centerline{\includegraphics[width=1.0\textwidth]{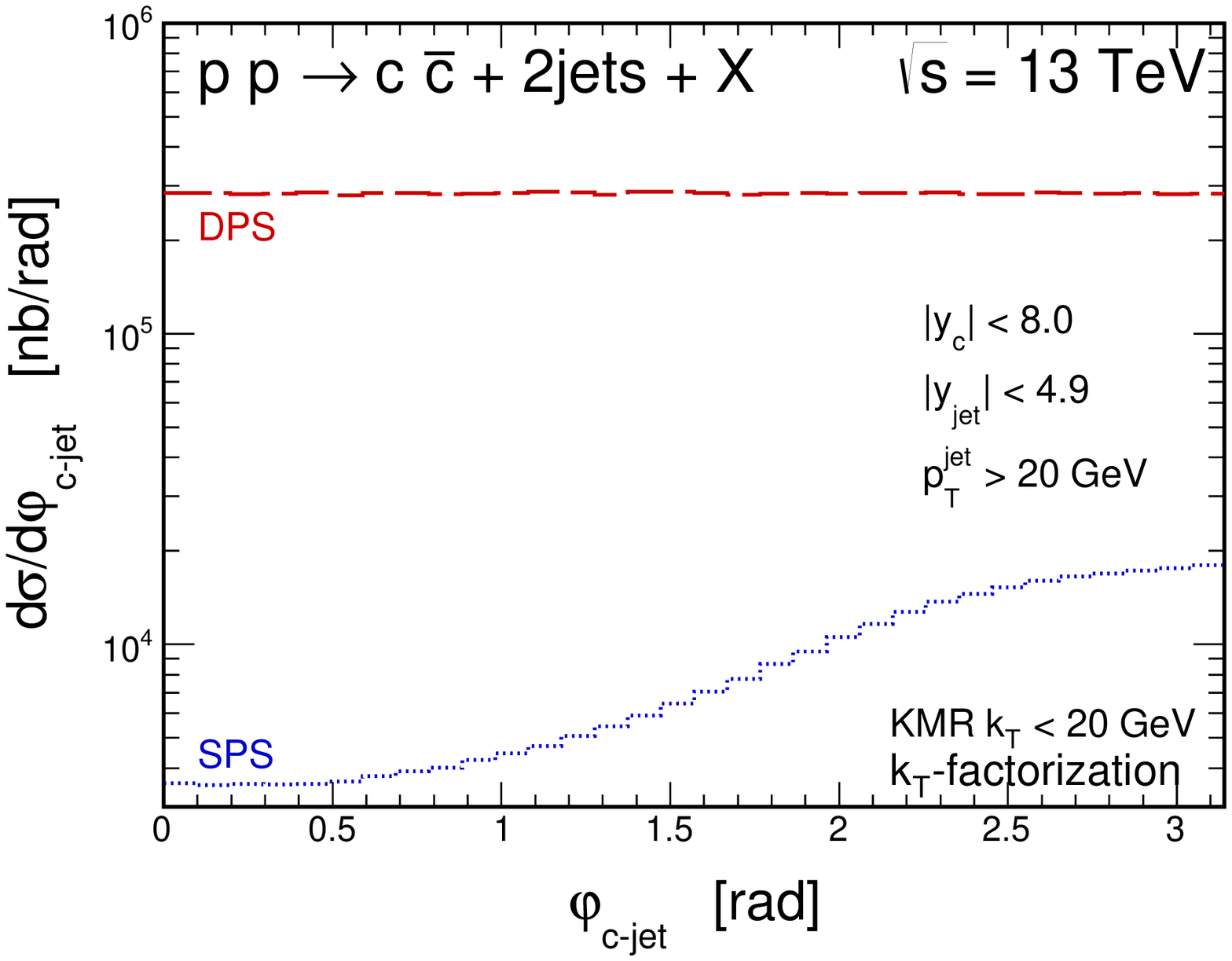}}
\end{minipage}
\hspace{0.5cm}
\begin{minipage}{0.47\textwidth}
 \centerline{\includegraphics[width=1.0\textwidth]{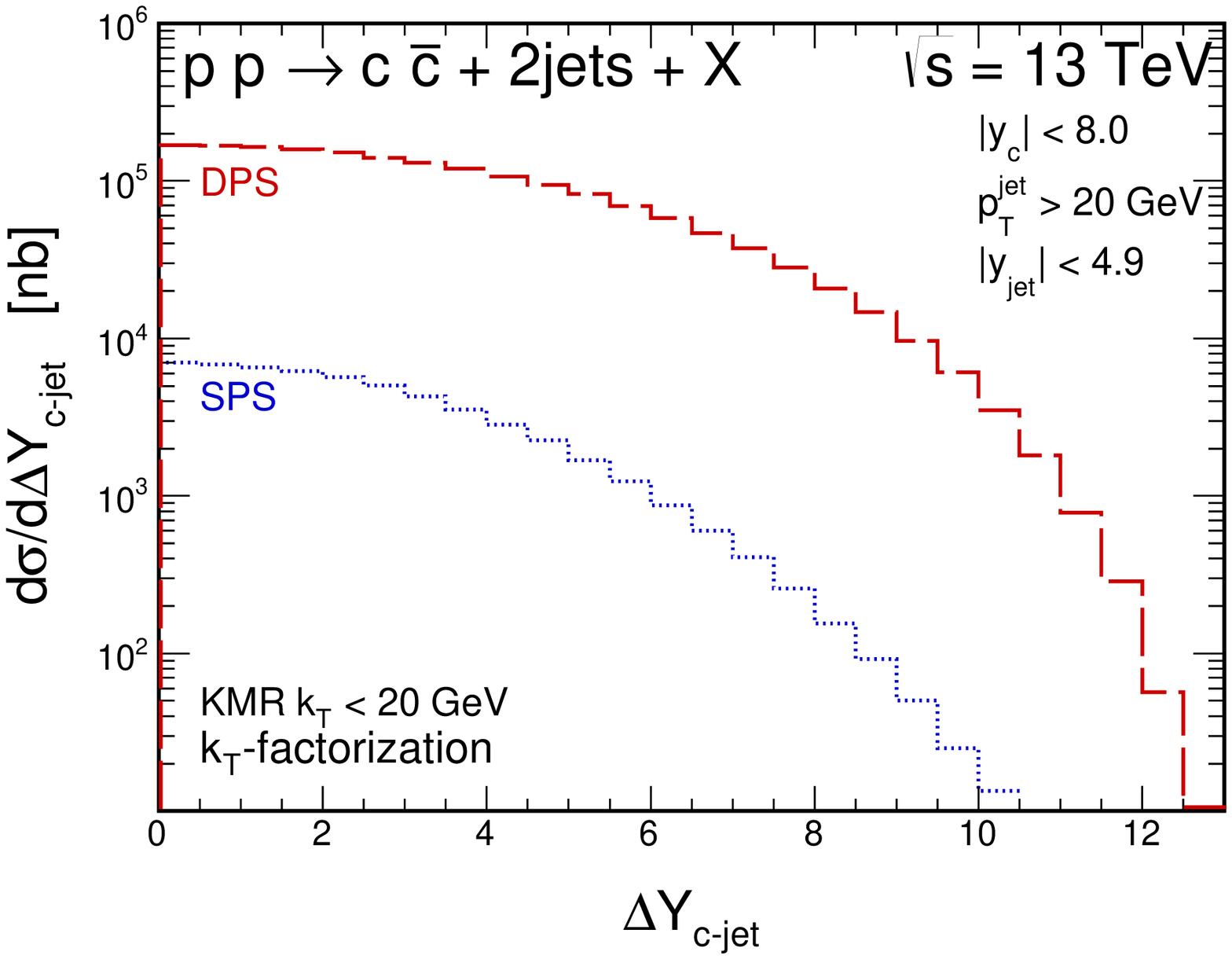}}
\end{minipage}
   \caption{
\small The same as in Fig.~\ref{fig:pt_jets} but for the azimuthal angle $\varphi_{c\mathrm{\textit{-jet}}}$ (left) and the rapidity difference $\Delta Y_{c\mathrm{\textit{-jet}}}$ (right) between the $c$-quark ($\bar{c}$-antiquark) and the leading jet.
 }
 \label{fig:corr_cjet}
\end{figure}

Considering the $c\bar c + \mathrm{2jets}$ final state one can also look at the correlations between $c$-quark and associated jet.
For example, in Fig.~\ref{fig:corr_cjet} we show the correlation distributions in the azimuthal angle $\varphi_{c\mathrm{\textit{-jet}}}$ (left panel)
and the rapidity difference $\Delta Y_{c\mathrm{\textit{-jet}}}$ between the $c$-quark ($\bar{c}$-antiquark) and the leading jet (right panel). 
Both correlation observables are predicted to be dominated by the DPS mechanism in the whole range of $\varphi_{c\mathrm{\textit{-jet}}}$ and $\Delta Y_{c\mathrm{\textit{-jet}}}$ , respectively.

%-----------------------------------------------------------------------------
%------------------------------------------------------------------------------

\subsection{$\bm{D^{0} + \mathrm{2jets}}$}

Now we move to the predictions for single $D^{0}$ meson production in association with exactly two jets.
The effects of the $c \to D^{0}$ hadronization are taken into account via standard fragmentation function technique.
For this purpose, we employ the scale-independent Peterson model of fragmentation function \cite{Peterson:1982ak} with $\varepsilon_{c} = 0.05$ which is commonly-used in the literature in the context of heavy quark fragmentation.  
Details of the fragmentation procedure applied here and useful discussion of the uncertainties related to the choice of the fragmentation function
can be found \textit{e.g.} in Ref.~\cite{Maciula:2013wg}. In the last step, the cross section for meson is normalized by the relevant branching fraction $\mathrm{BR}(c \to D^{0}) = 0.565$.    

In this analysis, the $D^{0}$ meson is required to have $|y^{D^{0}}| < 2.5$ and $p_{T}^{D^{0}} > 3.5$ GeV and the rapidities of both associated jets are $|y^{jet}| < 4.9$, which corresponds to the ATLAS detector acceptance. In Table~\ref{tab:cross sections_D} we collect the corresponding integrated cross sections for inclusive $D^{0}+\mathrm{2 jets}$ production in $pp$-scattering at $\sqrt{s} =$ 13 TeV for different cuts on transverse momenta of the associated jets, specified in the left column. The predictions are obtained within the $k_{T}$-factorization approach for the KMR uPDFs with the $k_{T} < p_{T,\mathrm{cut}}^{jet}$ constrain. We found large cross sections, of the order of a few, and up to even tens of microbarns, depending on the cuts on transverse momenta of the associated jets. The cross sections are dominated by the DPS mechanism with the relative DPS contribution
at the level of $70 - 80 \%$.  

%------------------------------------------------------------------------------------------------------------------------------
\begin{table}[tb]%
\caption{The calculated cross sections in microbarns for inclusive
  $D^{0}+\mathrm{2 jets}$ production in $pp$-scattering at $\sqrt{s} =$ 13 TeV for different cuts on transverse momenta of the associated jets. Here, the $D^{0}$ meson is required to have $|y^{D^{0}}| < 2.5$ and $p_{T}^{D^{0}} > 3.5$ GeV and the rapidities of the both associated jets are $|y^{jet}| < 4.9$, which corresponds to the ATLAS detector acceptance. The predictions were done within the $k_{T}$-factorization approach for the KMR uPDFs with the $k_{T} < p_{T,\mathrm{cut}}^{jet}$ constrain. }

\label{tab:cross sections_D}
\centering %
%\newcolumntype{Z}{>{\centering\arraybackslash}X}
%\newcommand{\tn}{\tabularnewline}
%\resizebox{\textwidth}{!}{%
\begin{tabularx}{1.\linewidth}{c c c c}
\\[-4.ex] 
\toprule[0.1em] %
\\[-4.ex] 
%\\[1.0ex]

\multirow{1}{7.5cm}{experimental jet-$p_{T}$ mode} & \multirow{1}{3.cm}{SPS} & \multirow{1}{3.cm}{DPS}  & \multirow{1}{3.cm}{$\frac{DPS}{SPS+DPS}$}  \\ [+0.1ex]
\bottomrule[0.1em]
\multirow{1}{7.5cm}{both jets $p_{T} > 20$ GeV} &                         \multirow{1}{3.cm}{3.74} & \multirow{1}{3.cm}{18.49} & \multirow{1}{3.cm}{$\;\;\;\;$83 \%}  \\ [-0.2ex]
\multirow{1}{7.5cm}{$p_{T}^{lead} > 35$ GeV, $\; p_{T}^{sub} > 20$ GeV} & \multirow{1}{3.cm}{1.76} & \multirow{1}{3.cm}{4.52} & \multirow{1}{3.cm}{$\;\;\;\;$72 \%} \\ [-0.2ex]
\multirow{1}{7.5cm}{$p_{T}^{lead} > 50$ GeV, $\; p_{T}^{sub} > 35$ GeV} & \multirow{1}{3.cm}{0.43} & \multirow{1}{3.cm}{1.25} & \multirow{1}{3.cm}{$\;\;\;\;$74 \%} \\ [-0.2ex]

\hline

\bottomrule[0.1em]

\end{tabularx}
%}
\end{table}
%-------------------------------------------------------------------------------------------------------------------------------

\begin{figure}[!h]
\begin{minipage}{0.47\textwidth}
 \centerline{\includegraphics[width=1.0\textwidth]{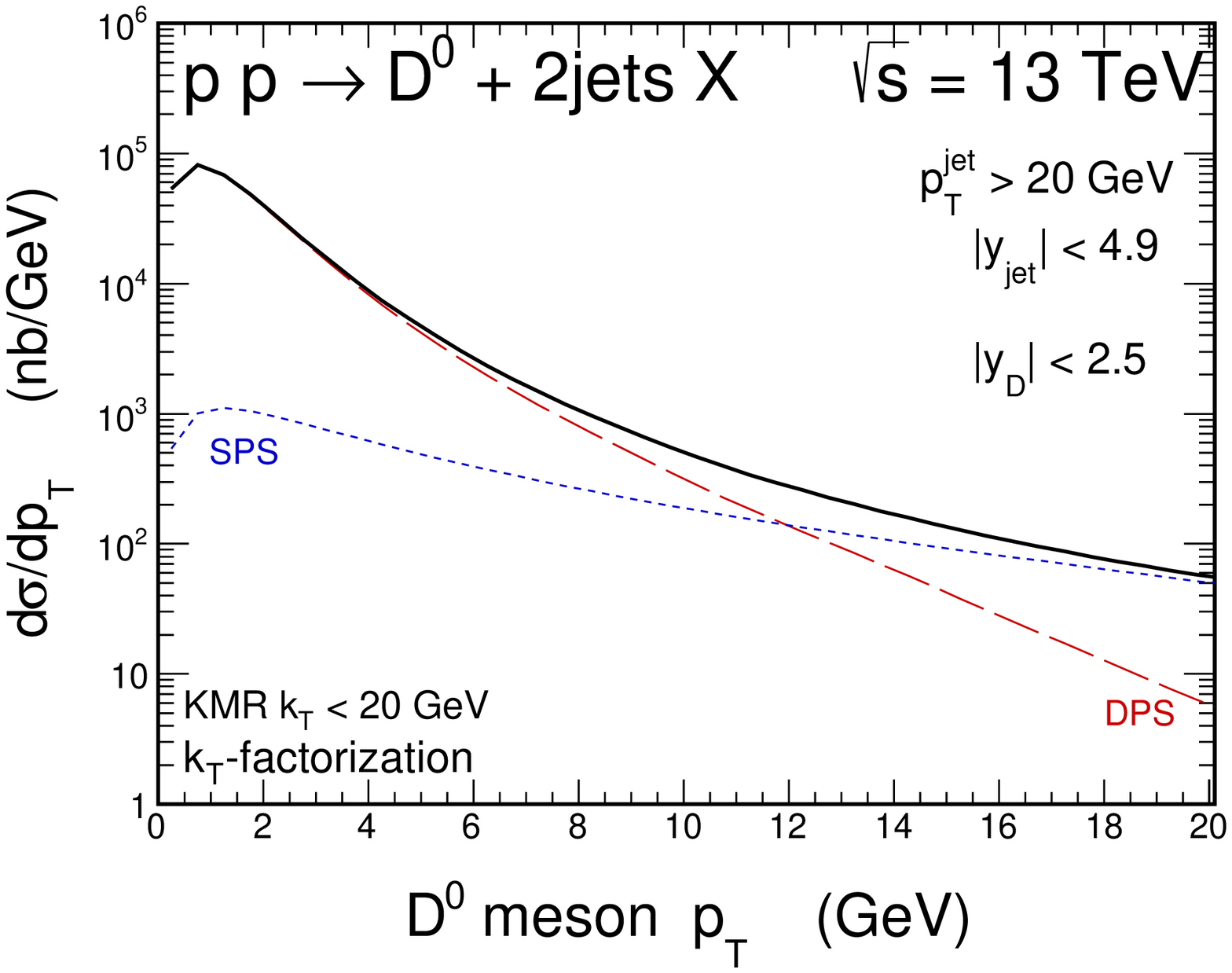}}
\end{minipage}
\hspace{0.5cm}
\begin{minipage}{0.47\textwidth}
 \centerline{\includegraphics[width=1.0\textwidth]{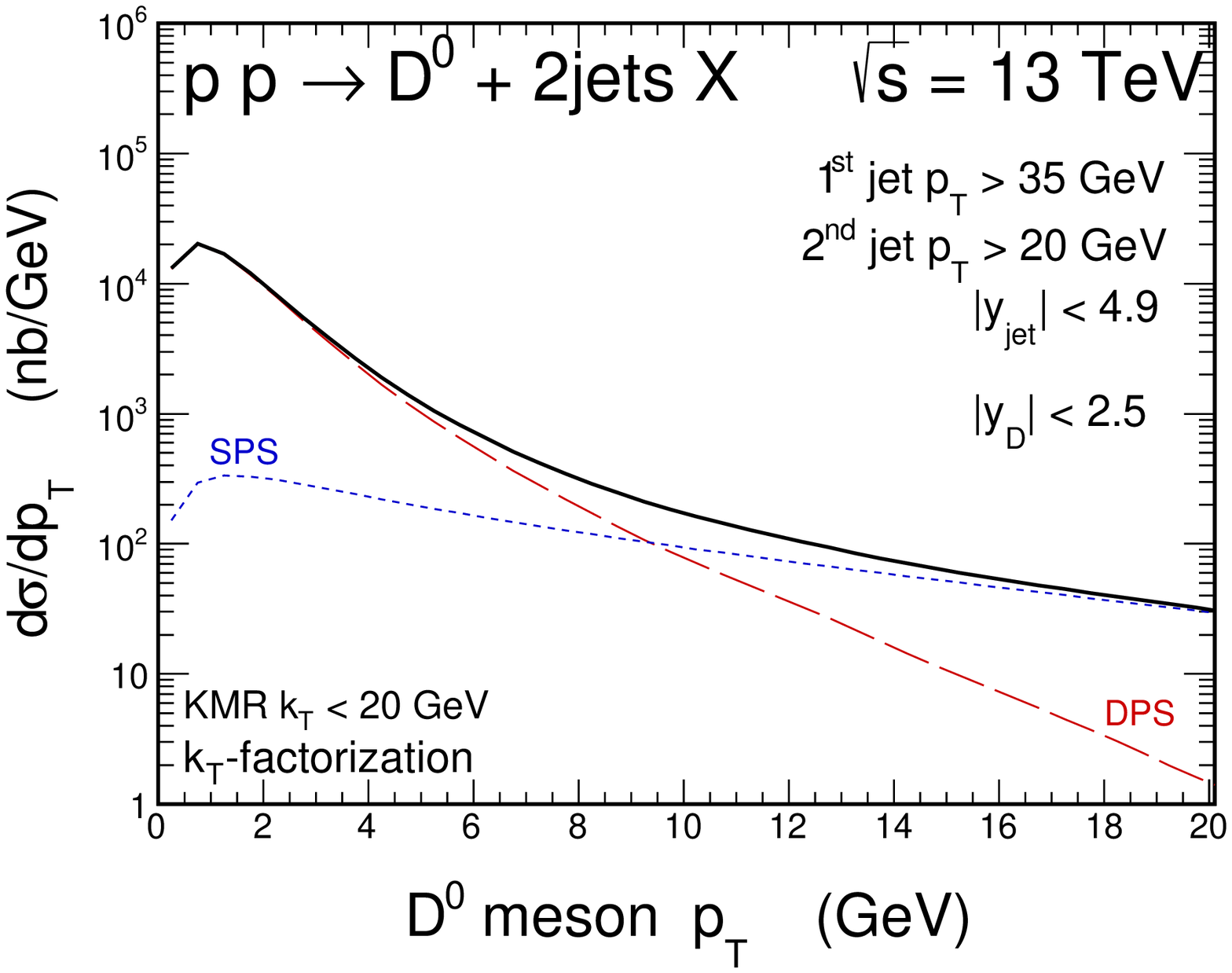}}
\end{minipage}
   \caption{
\small The transverse momentum distribution of the $D^{0}$ meson for SPS (dotted) and DPS (dashed) mechanisms for the ATLAS detector acceptance. The solid line represents a sum of the two components. The calculations are done within the $k_{T}$-factorization approach for the KMR uPDFs and with the $k_{T} < p_{T,\mathrm{cut}}^{jet}$ constrain. The left and right panels correspond to two different sets of cuts on the transverse momenta of the two associated jets. Details are specified in the figure.
 }
 \label{fig:ptD}
\end{figure}

In Fig.~\ref{fig:ptD} we show the differential cross section as a function of transverse momenta of the $D^{0}$ meson for two different sets of cuts on transverse momenta of the associated jets (left and right panel). The DPS (dashed line) and the SPS (dotted line) components are shown separately together with their sum (solid line). We observe that in the region of $D^{0}$ meson transverse momenta $p_{T} < 10$ GeV the DPS mechanism significantly dominates over the SPS one.

%-----------------------------------------------------------------------------
%------------------------------------------------------------------------------

\begin{figure}[!h]
\begin{minipage}{0.47\textwidth}
 \centerline{\includegraphics[width=1.0\textwidth]{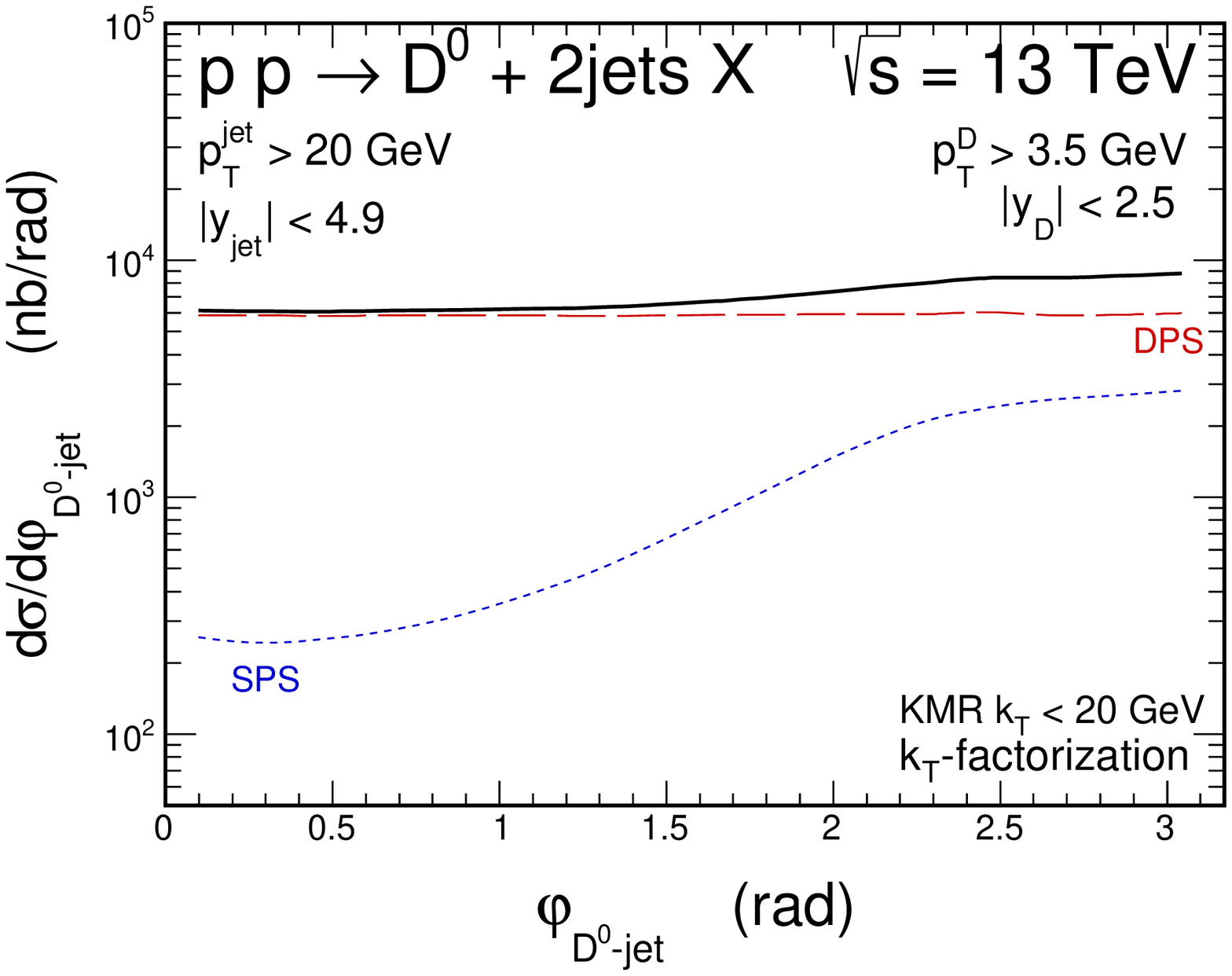}}
\end{minipage}
\hspace{0.5cm}
\begin{minipage}{0.47\textwidth}
 \centerline{\includegraphics[width=1.0\textwidth]{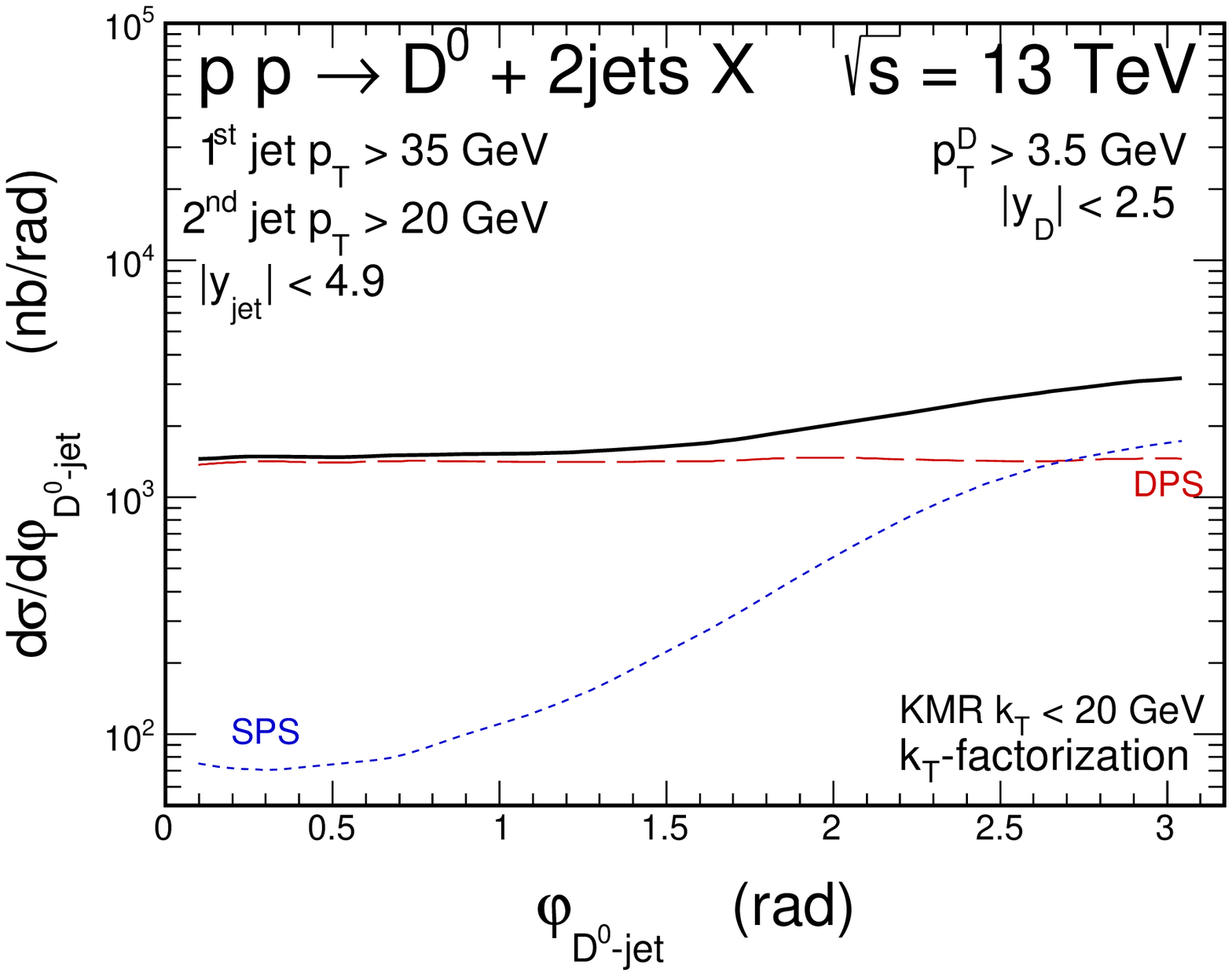}}
\end{minipage}
   \caption{
\small The same as in Fig.~\ref{fig:ptD} but for the azimuthal angle $\varphi_{D^{0}\mathrm{\textit{-jet}}}$
between the $D^{0}$ meson ($\overline{D^{0}}$ antimeson) and the leading jet in the case of inclusive $D^{0}+\mathrm{2 jets}$ production.}
 \label{fig:phiD-jet}
\end{figure}

Figure~\ref{fig:phiD-jet} shows a very interesting distributions in the azimuthal angle $\varphi_{D^{0}\mathrm{\textit{-jet}}}$
between the $D^{0}$ meson ($\overline{D^{0}}$ antimeson) and the leading jet, again for two different sets of cuts on transverse momenta of the associated jets (left and right panel). We see that the presence and the dominant role of the DPS component leads to a significant enhancement of the cross section and to a visible decorrelation of the distribution in contrast to the pure SPS-based predictions.   

%-----------------------------------------------------------------------------
%------------------------------------------------------------------------------

\subsection{$\bm{D^{0}\bar{D^{0}} + \mathrm{2jets}}$}

In the present analysis, we also consider the case of production of the $D^{0}\overline{D^{0}}$-pair in association with two jets.
So now both, $D^{0}$-meson and $\overline{D^{0}}$-antimeson are required to enter the ATLAS detector acceptance.
The corresponding theoretical cross sections are collected in Table~\ref{tab:cross sections_DD}. Here, the predicted cross sections for  $D^{0}\overline{D^{0}}+\mathrm{2 jets}$ are slightly
smaller than in the case of $D^{0}+\mathrm{2 jets}$ production (see Table~\ref{tab:cross sections_D}) but still large (in the best scenario, of the order of a few microbarns).
Also the relative DPS contribution is somewhat reduced and varies at the level of $50 - 70 \%$.   

%------------------------------------------------------------------------------------------------------------------------------
\begin{table}[tb]%
\caption{The same as in Table~\ref{tab:cross sections_D} but for inclusive $D^{0}\overline{D^{0}}+\mathrm{2 jets}$ production. Here both, $D^{0}$ meson and $\overline{D^{0}}$ antimeson are required to enter the ATLAS detector acceptance.}

\label{tab:cross sections_DD}
\centering %
%\newcolumntype{Z}{>{\centering\arraybackslash}X}
%\newcommand{\tn}{\tabularnewline}
%\resizebox{\textwidth}{!}{%
\begin{tabularx}{1.\linewidth}{c c c c}
\\[-4.ex] 
\toprule[0.1em] %
\\[-4.ex] 
%\\[1.0ex]

\multirow{1}{7.5cm}{experimental jet-$p_{T}$ mode} & \multirow{1}{3.cm}{SPS} & \multirow{1}{3.cm}{DPS}  & \multirow{1}{3.cm}{$\frac{DPS}{SPS+DPS}$}  \\ [+0.1ex]
\bottomrule[0.1em]
\multirow{1}{7.5cm}{both jets $p_{T} > 20$ GeV} &                         \multirow{1}{3.cm}{1.10} & \multirow{1}{3.cm}{2.35} & \multirow{1}{3.cm}{$\;\;\;\;$68 \%}  \\ [-0.2ex]
\multirow{1}{7.5cm}{$p_{T}^{lead} > 35$ GeV, $\; p_{T}^{sub} > 20$ GeV} & \multirow{1}{3.cm}{0.55} & \multirow{1}{3.cm}{0.58} & \multirow{1}{3.cm}{$\;\;\;\;$51 \%} \\ [-0.2ex]
\multirow{1}{7.5cm}{$p_{T}^{lead} > 50$ GeV, $\; p_{T}^{sub} > 35$ GeV} & \multirow{1}{3.cm}{0.15} & \multirow{1}{3.cm}{0.14} & \multirow{1}{3.cm}{$\;\;\;\;$52 \%} \\ [-0.2ex]

\hline

\bottomrule[0.1em]

\end{tabularx}
%}
\end{table}
%-------------------------------------------------------------------------------------------------------------------------------

\begin{figure}[!h]
\begin{minipage}{0.47\textwidth}
 \centerline{\includegraphics[width=1.0\textwidth]{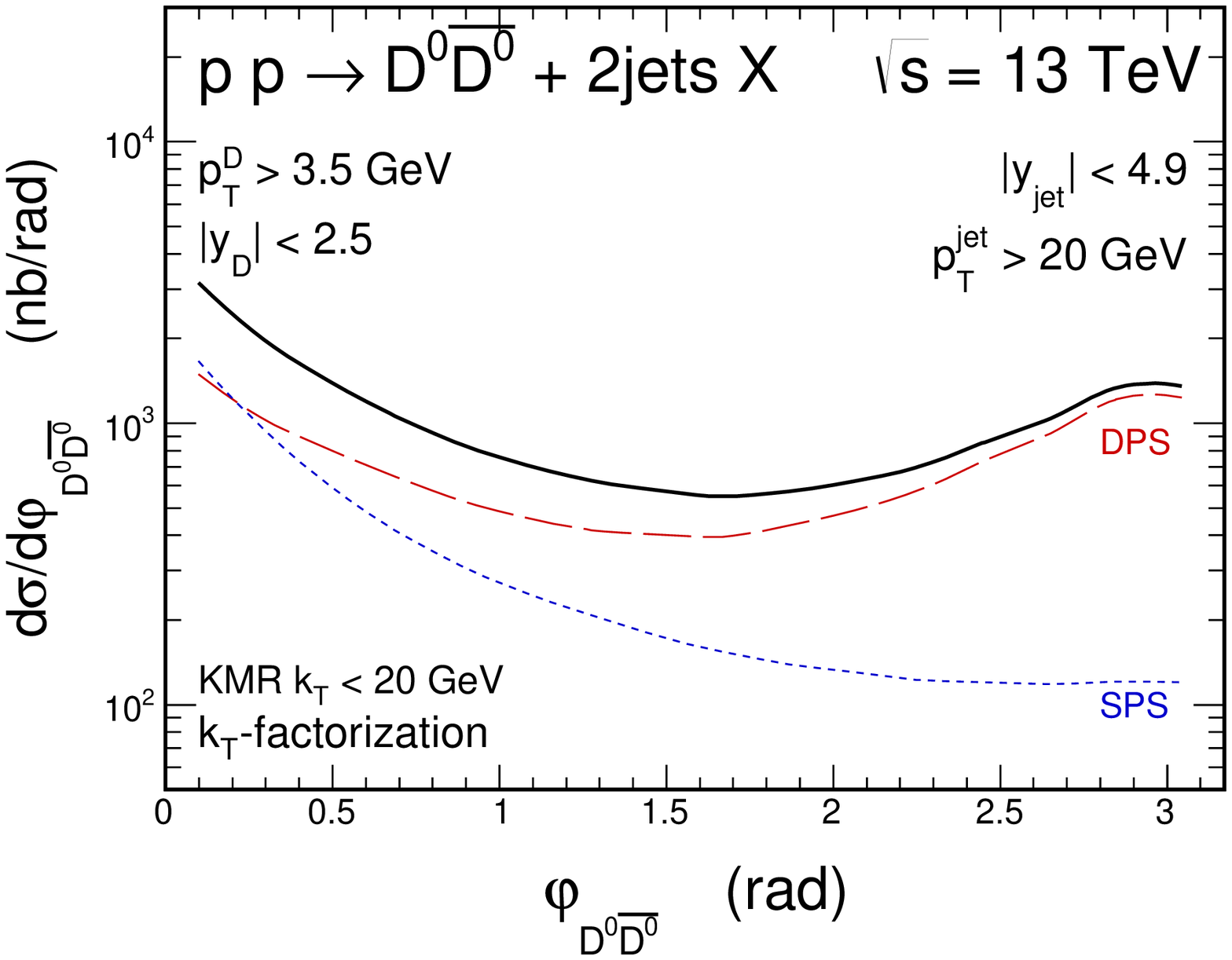}}
\end{minipage}
\hspace{0.5cm}
\begin{minipage}{0.47\textwidth}
 \centerline{\includegraphics[width=1.0\textwidth]{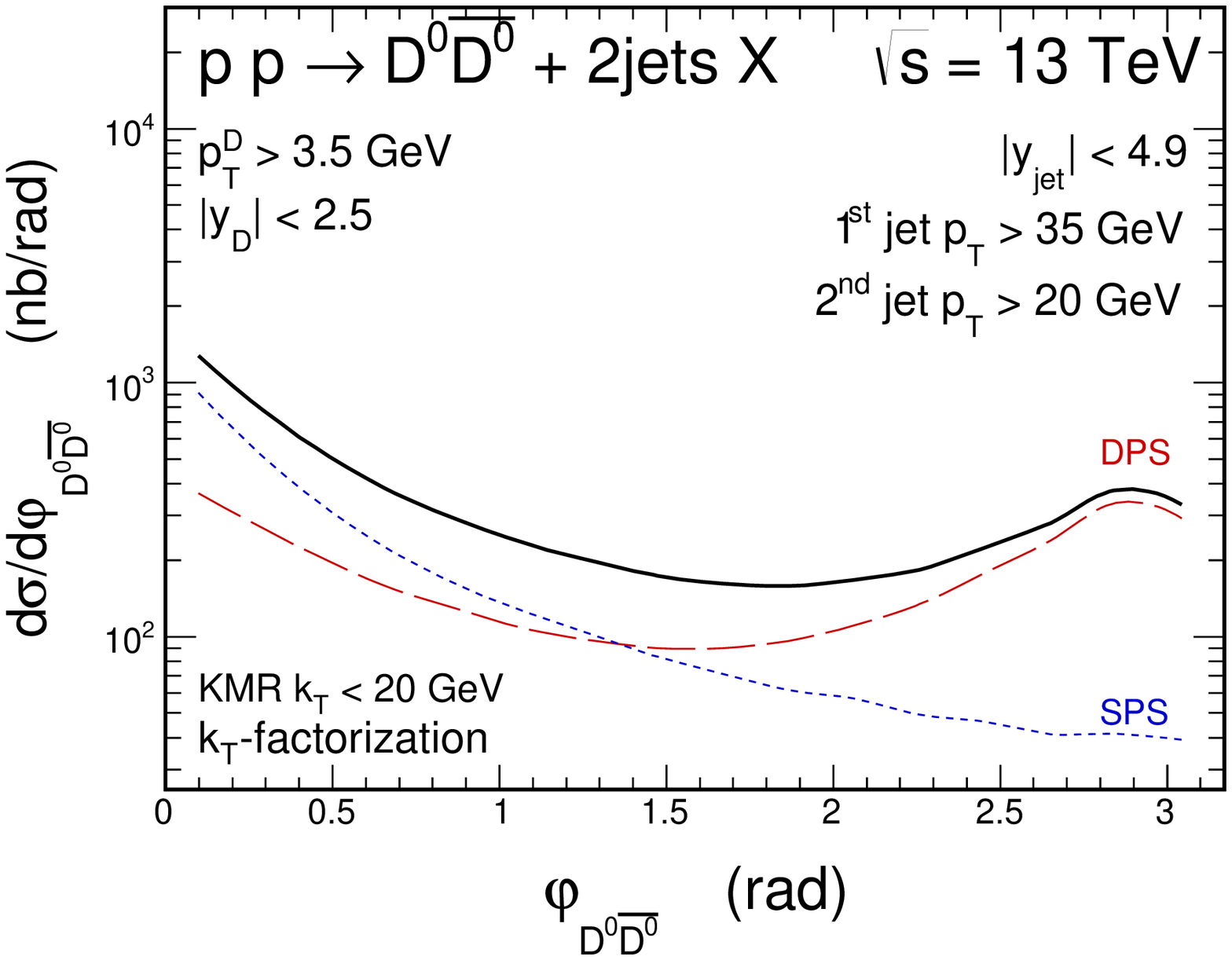}}
\end{minipage}
   \caption{
\small The same as in Fig.~\ref{fig:ptD} but for the azimuthal angle $\varphi_{D^{0}\overline{D^{0}}}$
between the $D^{0}$ meson and $\overline{D^{0}}$ antimeson in the case of inclusive $D^{0}\overline{D^{0}}+\mathrm{2 jets}$ production.}
 \label{fig:phiDDbar}
\end{figure}

In the case of the $D^{0}\overline{D^{0}}+\mathrm{2 jets}$ final state we also find a very interesting correlation observable that may be useful to distinguish between the DPS and SPS mechanisms. 
Figure~\ref{fig:phiDDbar} presents the distributions in azimuthal angle $\varphi_{D^{0}\overline{D^{0}}}$
between the $D^{0}$ meson and $\overline{D^{0}}$ antimeson in the case of $D^{0}\overline{D^{0}}+\mathrm{2 jets}$ production.
One can observe an evident enhancement of the cross section in the region of $\varphi_{D^{0}\overline{D^{0}}} > \frac{\pi}{2}$ caused by the presence of the DPS mechanism.

%-----------------------------------------------------------------------------
%------------------------------------------------------------------------------

%--------------------------
\section{Conclusions}
%--------------------------

In the present paper we have calculated for a first time cross sections 
for simultanous production of $c \bar c$ (or $D$ mesons) and dijets. 
Rather low transverse momentum cuts on jets have been used in order to 
enhance the DPS contribution. Both single and double parton scatttering
mechanisms have been included.

Several differential distributions have been shown and discussed.
The calculation have been performed for current LHC collision energy
$\sqrt{s}$ = 13 TeV within the $k_t$-factorization approach using the
Kimber-Martin-Ryskin unintegrated gluon distributions. The same
formalism turned out previously to be very successfull for description 
of production of one and two pairs of $c \bar c$ and production of dijets.

It was shown that the DPS contribution considerably dominates over the
SPS contribution for small transverse momenta of $c$ and/or $\bar c$.  
At larger transverse momenta of $c$ and/or $\bar c$ the SPS contribution
takes over.
The distribution in transverse momentum of $D$ mesons could be used to
pin down the competition of both mechanisms.
Very interesting are also azimuthal correlations between jets, between
jet and charm quark/antiquark or jet and $D$ mesons or even between charmed
mesons. The corresponding experimental distributions can be used
to further pin down underlying production mechanism.

We have identified regions of the phase space where the DPS
contributions significantly dominate. A future experimental cross sections in
these regions could be used to determine the $\sigma_{eff}$ parameter. 
In general, this quantity can dependend on several kinematical
quantities. Surprisingly similar values were obtained from the analysis of different
processes. We have discussed conditions how to extract $\sigma_{eff}$ 
from the discused in the present paper associated production of charm and dijets. 
Our analysis performed here shows that the extraction should be almost
free of systematic errors as the DPS contribution clearly dominantes
over the SPS one for the discussed reaction.

\vspace{1cm}

{\bf Acknowledgments}

We are particularly indebted to Andreas van Hameren for a help in using KaTie code.
This study was partially
supported by the Polish National Science Center grant
DEC-2014/15/B/ST2/02528 and by the Center for Innovation and
Transfer of Natural Sciences and Engineering Knowledge in
Rzesz{\'o}w.

%-------------------------------------------------------------------------------------


\begin{thebibliography}{100}
%-------------------------------------------------------------------------------------

%\cite{}
\bibitem{Luszczak:2011zp} 
  M.~Luszczak, R.~Maciu{\l}a and A.~Szczurek,
  %``Production of two $c \bar c$ pairs in double-parton scattering,''
  Phys.\ Rev.\ D {\bf 85}, 094034 (2012).
 % doi:10.1103/PhysRevD.85.094034
 % [arXiv:1111.3255 [hep-ph]].
  %%CITATION = doi:10.1103/PhysRevD.85.094034;%%
  %71 citations counted in INSPIRE as of 26 Jul 2017

%\cite{}
\bibitem{Maciula:2013kd} 
  R.~Maciu{\l}a and A.~Szczurek,
  %``Production of $c \bar c c \bar c$ in double-parton scattering within $k_{t}$-factorization approach -- meson-meson correlations,''
  Phys.\ Rev.\ D {\bf 87}, no. 7, 074039 (2013).
%  doi:10.1103/PhysRevD.87.074039
%  [arXiv:1301.4469 [hep-ph]].
  %%CITATION = doi:10.1103/PhysRevD.87.074039;%%
  %54 citations counted in INSPIRE as of 26 Jul 2017

%\cite{Cazaroto:2013fua}
\bibitem{Cazaroto:2013fua} 
  E.~R.~Cazaroto, V.~P.~Goncalves and F.~S.~Navarra,
  %``Heavy quark production and gluon saturation in double parton scattering at the LHC,''
  Phys.\ Rev.\ D {\bf 88}, no. 3, 034005 (2013).
 % doi:10.1103/PhysRevD.88.034005
 % [arXiv:1306.4169 [hep-ph]].
  %%CITATION = doi:10.1103/PhysRevD.88.034005;%%
  %18 citations counted in INSPIRE as of 26 Jul 2017

%\cite{}
\bibitem{vanHameren:2014ava} 
  A.~van Hameren, R.~Maciu{\l}a and A.~Szczurek,
  %``Single-parton scattering versus double-parton scattering in the production of two $c \bar c$ pairs and charmed meson correlations at the LHC,''
  Phys.\ Rev.\ D {\bf 89}, no. 9, 094019 (2014).
  %doi:10.1103/PhysRevD.89.094019
  %[arXiv:1402.6972 [hep-ph]].
  %%CITATION = doi:10.1103/PhysRevD.89.094019;%%
  %38 citations counted in INSPIRE as of 26 Jul 2017

%\cite{Maciula:2016wci}
\bibitem{Maciula:2016wci} 
  R.~Maciu{\l}a, V.~A.~Saleev, A.~V.~Shipilova and A.~Szczurek,
  %``New mechanisms for double charmed meson production at the LHCb,''
  Phys.\ Lett.\ B {\bf 758}, 458 (2016).
 % doi:10.1016/j.physletb.2016.05.052
 % [arXiv:1601.06981 [hep-ph]].
  %%CITATION = doi:10.1016/j.physletb.2016.05.052;%%
  %15 citations counted in INSPIRE as of 26 Jul 2017

%\cite{Maciula:2017meb}
\bibitem{Maciula:2017meb} 
  R.~Maciu{\l}a and A.~Szczurek,
  %``Can the triple-parton scattering be observed in open charm meson production at the LHC?,''
  arXiv:1703.07163 [hep-ph].
  %%CITATION = ARXIV:1703.07163;%%

%\cite{Maciula:2016kkx}
\bibitem{Maciula:2016kkx} 
  R.~Maciu{\l}a and A.~Szczurek,
  %``Charm quark and meson production in association with single-jet at the LHC,''
  Phys.\ Rev.\ D {\bf 94}, no. 11, 114037 (2016).
 % doi:10.1103/PhysRevD.94.114037
 % [arXiv:1610.01810 [hep-ph]].
  %%CITATION = doi:10.1103/PhysRevD.94.114037;%%
  
%\cite{}
\bibitem{Abazov:2014qba} 
  V.~M.~Abazov {\it et al.} [D0 Collaboration],
  %``Observation and studies of double $J/\psi$ production at the Tevatron,''
  Phys.\ Rev.\ D {\bf 90}, no. 11, 111101 (2014).
%  doi:10.1103/PhysRevD.90.111101
%  [arXiv:1406.2380 [hep-ex]].
  %%CITATION = doi:10.1103/PhysRevD.90.111101;%%
  %49 citations counted in INSPIRE as of 26 Jul 2017
  
%\cite{}
\bibitem{Khachatryan:2014iia} 
  V.~Khachatryan {\it et al.} [CMS Collaboration],
  %``Measurement of prompt $J/\psi$ pair production in pp collisions at $ \sqrt{s} $ = 7 Tev,''
  J. High Energy Phys. {\bf 09}, 094 (2014).
  %doi:10.1007/JHEP09(2014)094
  %[arXiv:1406.0484 [hep-ex]].
  %%CITATION = doi:10.1007/JHEP09(2014)094;%%
  %49 citations counted in INSPIRE as of 26 Jul 2017  
  
%\cite{Aaboud:2016fzt}
\bibitem{Aaboud:2016fzt} 
  M.~Aaboud {\it et al.} [ATLAS Collaboration],
  %``Measurement of the prompt J/ $\psi $ pair production cross-section in pp collisions at $\sqrt{s} = 8$  TeV with the ATLAS detector,''
  Eur.\ Phys.\ J.\ C {\bf 77}, no. 2, 76 (2017).
  %doi:10.1140/epjc/s10052-017-4644-9
  %[arXiv:1612.02950 [hep-ex]].
  %%CITATION = doi:10.1140/epjc/s10052-017-4644-9;%%
  %7 citations counted in INSPIRE as of 26 Jul 2017  


%\cite{Szczurek:2017uvc}
\bibitem{Szczurek:2017uvc} 
  A.~Szczurek, A.~Cisek and W.~Sch\"afer,
  %``Some new aspects of quarkonia production at the LHC,''
  Acta Phys.\ Polon.\ B {\bf 48}, 1207 (2017)
%  doi:10.5506/APhysPolB.48.1207
  [arXiv:1704.00444 [hep-ph]].
  %%CITATION = doi:10.5506/APhysPolB.48.1207;%%

\bibitem{CSSB2017}
A. Cisek, W. Sch\"afer, A. Szczurek and S. Baranov, a paper in
preparation.
  

%\cite{Catani:1990eg}
\bibitem{Catani:1990eg} 
  S.~Catani, M.~Ciafaloni and F.~Hautmann,
  %``High-energy factorization and small x heavy flavor production,''
  Nucl.\ Phys.\ B {\bf 366}, 135 (1991).
%  doi:10.1016/0550-3213(91)90055-3
  %%CITATION = doi:10.1016/0550-3213(91)90055-3;%%
  %930 citations counted in INSPIRE as of 22 Sep 2016

%\cite{Nefedov:2013ywa}
\bibitem{Nefedov:2013ywa} 
  M.~A.~Nefedov, V.~A.~Saleev and A.~V.~Shipilova,
  %``Dijet azimuthal decorrelations at the LHC in the parton Reggeization approach,''
  Phys.\ Rev.\ D {\bf 87}, no. 9, 094030 (2013)
  %doi:10.1103/PhysRevD.87.094030
  [arXiv:1304.3549 [hep-ph]].
  %%CITATION = doi:10.1103/PhysRevD.87.094030;%%
  %32 citations counted in INSPIRE as of 22 Sep 2016

%\cite{Nefedov:2012cq}
\bibitem{Nefedov:2012cq} 
  M.~A.~Nefedov, N.~N.~Nikolaev and V.~A.~Saleev,
  %``Drell-Yan lepton pair production at high energies in the Parton Reggeization Approach,''
  Phys.\ Rev.\ D {\bf 87}, no. 1, 014022 (2013)
%  doi:10.1103/PhysRevD.87.014022
  [arXiv:1211.5539 [hep-ph]].
  %%CITATION = doi:10.1103/PhysRevD.87.014022;%%
  %19 citations counted in INSPIRE as of 22 Sep 2016

%\cite{Nefedov:2015ara}
\bibitem{Nefedov:2015ara} 
  M.~Nefedov and V.~Saleev,
  %``Diphoton production at the Tevatron and the LHC in the NLO approximation of the parton Reggeization approach,''
  Phys.\ Rev.\ D {\bf 92}, no. 9, 094033 (2015)
%  doi:10.1103/PhysRevD.92.094033
  [arXiv:1505.01718 [hep-ph]].
  %%CITATION = doi:10.1103/PhysRevD.92.094033;%%
  %2 citations counted in INSPIRE as of 22 Sep 2016
  
%\cite{Nefedov:2016clr}
\bibitem{Nefedov:2016clr} 
  M.~Nefedov and V.~Saleev,
  %``Towards NLO calculations in the parton Reggeization approach,''
  arXiv:1608.04201 [hep-ph].
  %%CITATION = ARXIV:1608.04201;%%
  
%\cite{vanHameren:2012if}
\bibitem{vanHameren:2012if} 
  A.~van Hameren, P.~Kotko and K.~Kutak,
  %``Helicity amplitudes for high-energy scattering,''
  J. High Energy Phys {\bf 01}, 078 (2013)
  %doi:10.1007/JHEP01(2013)078
  [arXiv:1211.0961 [hep-ph]].
  %%CITATION = doi:10.1007/JHEP01(2013)078;%%
  %51 citations counted in INSPIRE as of 22 Sep 2016

\bibitem{vanHameren:2014iua} 
  A.~van Hameren,
  %``BCFW recursion for off-shell gluons,''
  JHEP {\bf 1407}, 138 (2014)
%  doi:10.1007/JHEP07(2014)138
  [arXiv:1404.7818 [hep-ph]].
  %%CITATION = doi:10.1007/JHEP07(2014)138;%%
  %26 citations counted in INSPIRE as of 22 Sep 2016  

%\cite{Bury:2015dla}
\bibitem{Bury:2015dla} 
  M.~Bury and A.~van Hameren,
  %``Numerical evaluation of multi-gluon amplitudes for High Energy Factorization,''
  Comput.\ Phys.\ Commun.\  {\bf 196}, 592 (2015)
 % doi:10.1016/j.cpc.2015.06.023
  [arXiv:1503.08612 [hep-ph]].
  %%CITATION = doi:10.1016/j.cpc.2015.06.023;%%
  %18 citations counted in INSPIRE as of 22 Sep 2016

%\cite{vanHameren:2015wva}
\bibitem{vanHameren:2015wva} 
  A.~van Hameren, R.~Maciu{\l}a and A.~Szczurek,
  %``Production of two charm quark-antiquark pairs in single-parton scattering within the $k_t$-factorization approach,''
  Phys.\ Lett.\ B {\bf 748}, 167 (2015)
%  doi:10.1016/j.physletb.2015.06.061
  [arXiv:1504.06490 [hep-ph]].
  %%CITATION = doi:10.1016/j.physletb.2015.06.061;%%
  %11 citations counted in INSPIRE as of 22 Sep 2016
  
  %\cite{Kutak:2016mik}
\bibitem{Kutak:2016mik} 
  K.~Kutak, R.~Maciu{\l}a, M.~Serino, A.~Szczurek and A.~van Hameren,
  %``Four-jet production in single- and double-parton scattering within high-energy factorization,''
  J. High Energy Phys. {\bf 04}, 175 (2016)
%  doi:10.1007/JHEP04(2016)175
  [arXiv:1602.06814 [hep-ph]].
  %%CITATION = doi:10.1007/JHEP04(2016)175;%%
  %9 citations counted in INSPIRE as of 21 Sep 2016  
  
%\cite{vanHameren:2016kkz}
\bibitem{vanHameren:2016kkz} 
  A.~van Hameren,
  %``KaTie: for parton-level event generation with k_T-dependent initial states,''
  arXiv:1611.00680 [hep-ph].
  %%CITATION = ARXIV:1611.00680;%%
  %3 citations counted in INSPIRE as of 03 Jun 2017  

%\cite{Caravaglios:1995cd}
\bibitem{Caravaglios:1995cd} 
  F.~Caravaglios and M.~Moretti,
  %``An algorithm to compute Born scattering amplitudes without Feynman graphs,''
  Phys.\ Lett.\ B {\bf 358}, 332 (1995)
%  doi:10.1016/0370-2693(95)00971-M
  [hep-ph/9507237].
  %%CITATION = doi:10.1016/0370-2693(95)00971-M;%%
  %180 citations counted in INSPIRE as of 03 Jun 2017

%\cite{vanHameren:2007pt}
\bibitem{vanHameren:2007pt} 
  A.~van Hameren,
  %``PARNI for importance sampling and density estimation,''
  Acta Phys.\ Polon.\ B {\bf 40}, 259 (2009)
  [arXiv:0710.2448 [hep-ph]].
  %%CITATION = ARXIV:0710.2448;%%
  %31 citations counted in INSPIRE as of 03 Jun 2017
  
%\cite{vanHameren:2010gg}
\bibitem{vanHameren:2010gg} 
  A.~van Hameren,
  %``Kaleu: A General-Purpose Parton-Level Phase Space Generator,''
  arXiv:1003.4953 [hep-ph].
  %%CITATION = ARXIV:1003.4953;%%
  %45 citations counted in INSPIRE as of 03 Jun 2017

%\cite{Kimber:2001sc}
\bibitem{Kimber:2001sc} 
  M.~A.~Kimber, A.~D.~Martin and M.~G.~Ryskin,
  %``Unintegrated parton distributions,''
  Phys.\ Rev.\ D {\bf 63}, 114027 (2001)
%  doi:10.1103/PhysRevD.63.114027
  [hep-ph/0101348].
  %%CITATION = doi:10.1103/PhysRevD.63.114027;%%
  %277 citations counted in INSPIRE as of 23 Sep 2016
  
 %\cite{Watt:2003vf}
\bibitem{Watt:2003vf}
  G.~Watt, A.~D.~Martin and M.~G.~Ryskin,
  %``Unintegrated parton distributions and electroweak boson production at hadron colliders,''
  Phys.\ Rev.\ D {\bf 70} (2004) 014012
   Erratum: [Phys.\ Rev.\ D {\bf 70} (2004) 079902]
%  doi:10.1103/PhysRevD.70.014012, 10.1103/PhysRevD.70.079902
  [hep-ph/0309096].
  %%CITATION = doi:10.1103/PhysRevD.70.014012, 10.1103/PhysRevD.70.079902;%%
  %82 citations counted in INSPIRE as of 23 Sep 2016 

%\cite{Harland-Lang:2014zoa}
\bibitem{Harland-Lang:2014zoa} 
  L.~A.~Harland-Lang, A.~D.~Martin, P.~Motylinski and R.~S.~Thorne,
  %``Parton distributions in the LHC era: MMHT 2014 PDFs,''
  Eur.\ Phys.\ J.\ C {\bf 75}, no. 5, 204 (2015)
%  doi:10.1140/epjc/s10052-015-3397-6
  [arXiv:1412.3989 [hep-ph]].
  %%CITATION = doi:10.1140/epjc/s10052-015-3397-6;%%
  %367 citations counted in INSPIRE as of 03 Jun 2017  

%\cite{Diehl:2011tt}
\bibitem{Diehl:2011tt}
  M.~Diehl and A.~Schafer,
  %``Theoretical considerations on multiparton interactions in QCD,''
  Phys.\ Lett.\ B {\bf 698} (2011) 389
  %doi:10.1016/j.physletb.2011.03.024
  [arXiv:1102.3081 [hep-ph]].
  %%CITATION = doi:10.1016/j.physletb.2011.03.024;%%
  %104 citations counted in INSPIRE as of 06 Jun 2017
  
%\cite{Diehl:2011yj}
\bibitem{Diehl:2011yj} 
  M.~Diehl, D.~Ostermeier and A.~Schafer,
  %``Elements of a theory for multiparton interactions in QCD,''
  JHEP {\bf 1203}, 089 (2012)
  Erratum: [JHEP {\bf 1603}, 001 (2016)]
  %doi:10.1007/JHEP03(2012)089, 10.1007/JHEP03(2016)001
  [arXiv:1111.0910 [hep-ph]].
  %%CITATION = doi:10.1007/JHEP03(2012)089, 10.1007/JHEP03(2016)001;%%
  %142 citations counted in INSPIRE as of 06 Jun 2017


%\cite{Gaunt:2009re}
\bibitem{Gaunt:2009re} 
  J.~R.~Gaunt and W.~J.~Stirling,
  %``Double Parton Distributions Incorporating Perturbative QCD Evolution and Momentum and Quark Number Sum Rules,''
  JHEP {\bf 1003}, 005 (2010)
 % doi:10.1007/JHEP03(2010)005
  [arXiv:0910.4347 [hep-ph]].
  %%CITATION = doi:10.1007/JHEP03(2010)005;%%
  %128 citations counted in INSPIRE as of 06 Jun 2017
    
%\cite{Abe:1997bp}
\bibitem{Abe:1997bp} 
  F.~Abe {\it et al.} [CDF Collaboration],
  %``Measurement of double parton scattering in $\bar{p}p$ collisions at $\sqrt{s} = 1.8$ TeV,''
  Phys.\ Rev.\ Lett.\  {\bf 79}, 584 (1997).
 % doi:10.1103/PhysRevLett.79.584
  %%CITATION = doi:10.1103/PhysRevLett.79.584;%%
  %152 citations counted in INSPIRE as of 06 Jun 2017

%\cite{Abe:1997xk}
\bibitem{Abe:1997xk} 
  F.~Abe {\it et al.} [CDF Collaboration],
  %``Double parton scattering in $\bar{p}p$ collisions at $\sqrt{s} = 1.8 $TeV,''
  Phys.\ Rev.\ D {\bf 56}, 3811 (1997).
%  doi:10.1103/PhysRevD.56.3811
  %%CITATION = doi:10.1103/PhysRevD.56.3811;%%
  %331 citations counted in INSPIRE as of 06 Jun 2017

%\cite{Abazov:2009gc}
\bibitem{Abazov:2009gc} 
  V.~M.~Abazov {\it et al.} [D0 Collaboration],
  %``Double parton interactions in $\gamma$+3 jet events in $p p^-$ bar collisions $\sqrt{s}=1.96$ TeV.,''
  Phys.\ Rev.\ D {\bf 81}, 052012 (2010).
%  doi:10.1103/PhysRevD.81.052012
 % [arXiv:0912.5104 [hep-ex]].
  %%CITATION = doi:10.1103/PhysRevD.81.052012;%%
  %192 citations counted in INSPIRE as of 06 Jun 2017

%\cite{Aaij:2012dz}
\bibitem{Aaij:2012dz}
  R.~Aaij {\it et al.} [LHCb Collaboration],
  %``Observation of double charm production involving open charm in pp collisions at $\sqrt{s}$ = 7 TeV,''
  JHEP {\bf 1206} (2012) 141
   Addendum: [JHEP {\bf 1403} (2014) 108]
%  doi:10.1007/JHEP03(2014)108, 10.1007/JHEP06(2012)141
  [arXiv:1205.0975 [hep-ex]].
  %%CITATION = doi:10.1007/JHEP03(2014)108, 10.1007/JHEP06(2012)141;%%
  %144 citations counted in INSPIRE as of 06 Jun 2017

%\cite{Aad:2013bjm}
\bibitem{Aad:2013bjm} 
  G.~Aad {\it et al.} [ATLAS Collaboration],
  %``Measurement of hard double-parton interactions in $W(\to l\nu)$+ 2 jet events at $\sqrt{s}$=7 TeV with the ATLAS detector,''
  New J.\ Phys.\  {\bf 15}, 033038 (2013)
 % doi:10.1088/1367-2630/15/3/033038
  [arXiv:1301.6872 [hep-ex]].
  %%CITATION = doi:10.1088/1367-2630/15/3/033038;%%
  %179 citations counted in INSPIRE as of 06 Jun 2017
  
%\cite{Chatrchyan:2013xxa}
\bibitem{Chatrchyan:2013xxa} 
  S.~Chatrchyan {\it et al.} [CMS Collaboration],
  %``Study of double parton scattering using W + 2-jet events in proton-proton collisions at $\sqrt{s}$ = 7 TeV,''
  JHEP {\bf 1403}, 032 (2014)
%  doi:10.1007/JHEP03(2014)032
  [arXiv:1312.5729 [hep-ex]].
  %%CITATION = doi:10.1007/JHEP03(2014)032;%%
  %104 citations counted in INSPIRE as of 06 Jun 2017

%\cite{Aad:2014rua}
\bibitem{Aad:2014rua} 
  G.~Aad {\it et al.} [ATLAS Collaboration],
  %``Measurement of the production cross section of prompt $J/\psi$ mesons in association with a $W^\pm$ boson in $pp$ collisions at $\sqrt{s} =$ 7 TeV with the ATLAS detector,''
  JHEP {\bf 1404}, 172 (2014)
%  doi:10.1007/JHEP04(2014)172
  [arXiv:1401.2831 [hep-ex]].
  %%CITATION = doi:10.1007/JHEP04(2014)172;%%
  %56 citations counted in INSPIRE as of 06 Jun 2017
  
%\cite{Aaboud:2016dea}
\bibitem{Aaboud:2016dea} 
  M.~Aaboud {\it et al.} [ATLAS Collaboration],
  %``Study of hard double-parton scattering in four-jet events in pp collisions at $ \sqrt{s}=7 $ TeV with the ATLAS experiment,''
  JHEP {\bf 1611}, 110 (2016)
 % doi:10.1007/JHEP11(2016)110
  [arXiv:1608.01857 [hep-ex]].
  %%CITATION = doi:10.1007/JHEP11(2016)110;%%
  %10 citations counted in INSPIRE as of 06 Jun 2017


%\cite{Echevarria:2015ufa}
\bibitem{Echevarria:2015ufa} 
  M.~G.~Echevarria, T.~Kasemets, P.~J.~Mulders and C.~Pisano,
  %``Polarization effects in double open-charm production at LHCb,''
  JHEP {\bf 1504}, 034 (2015)
  %doi:10.1007/JHEP04(2015)034
  [arXiv:1501.07291 [hep-ph]].
  %%CITATION = doi:10.1007/JHEP04(2015)034;%%
  %18 citations counted in INSPIRE as of 06 Jun 2017

%\cite{Ryskin:2011kk}
\bibitem{Ryskin:2011kk} 
  M.~G.~Ryskin and A.~M.~Snigirev,
  %``A Fresh look at double parton scattering,''
  Phys.\ Rev.\ D {\bf 83}, 114047 (2011)
  %doi:10.1103/PhysRevD.83.114047
  [arXiv:1103.3495 [hep-ph]].
  %%CITATION = doi:10.1103/PhysRevD.83.114047;%%
  %70 citations counted in INSPIRE as of 06 Jun 2017

%\cite{Gaunt:2012dd}
\bibitem{Gaunt:2012dd} 
  J.~R.~Gaunt,
  %``Single Perturbative Splitting Diagrams in Double Parton Scattering,''
  JHEP {\bf 1301}, 042 (2013)
  %doi:10.1007/JHEP01(2013)042
  [arXiv:1207.0480 [hep-ph]].
  %%CITATION = doi:10.1007/JHEP01(2013)042;%%
  %43 citations counted in INSPIRE as of 06 Jun 2017

%\cite{Gaunt:2014rua}
\bibitem{Gaunt:2014rua} 
  J.~R.~Gaunt, R.~Maciu{\l}a and A.~Szczurek,
  %``Conventional versus single-ladder-splitting contributions to double parton scattering production of two quarkonia, two Higgs bosons and $c \bar c c \bar c$,''
  Phys.\ Rev.\ D {\bf 90}, no. 5, 054017 (2014)
 % doi:10.1103/PhysRevD.90.054017
  [arXiv:1407.5821 [hep-ph]].
  %%CITATION = doi:10.1103/PhysRevD.90.054017;%%
  %40 citations counted in INSPIRE as of 06 Jun 2017  

%\cite{Golec-Biernat:2015aza}
\bibitem{Golec-Biernat:2015aza} 
  K.~Golec-Biernat, E.~Lewandowska, M.~Serino, Z.~Snyder and A.~M.~Stasto,
  %``Constraining the double gluon distribution by the single gluon distribution,''
  Phys.\ Lett.\ B {\bf 750}, 559 (2015)
 % doi:10.1016/j.physletb.2015.09.067
  [arXiv:1507.08583 [hep-ph]].
  %%CITATION = doi:10.1016/j.physletb.2015.09.067;%%
  %14 citations counted in INSPIRE as of 06 Jun 2017

%\cite{Maciula:2013wg}
\bibitem{Maciula:2013wg} 
  R.~Maciu{\l}a and A.~Szczurek,
  %``Open charm production at the LHC - $k_{t}$-factorization approach,''
  Phys.\ Rev.\ D {\bf 87}, no. 9, 094022 (2013)
 % doi:10.1103/PhysRevD.87.094022
  [arXiv:1301.3033 [hep-ph]].
  %%CITATION = doi:10.1103/PhysRevD.87.094022;%%
  %79 citations counted in INSPIRE as of 06 Jun 2017

%\cite{Karpishkov:2016hnx}
\bibitem{Karpishkov:2016hnx} 
  A.~Karpishkov, V.~Saleev and A.~Shipilova,
  %``Large-$p_T$ production of D mesons at the LHCb in the parton Reggeization approach,''
  Phys.\ Rev.\ D {\bf 94}, no. 11, 114012 (2016)
%  doi:10.1103/PhysRevD.94.114012
  [arXiv:1610.04975 [hep-ph]].
  %%CITATION = doi:10.1103/PhysRevD.94.114012;%%
  %2 citations counted in INSPIRE as of 06 Jun 2017

%\cite{Peterson:1982ak}
\bibitem{Peterson:1982ak} 
  C.~Peterson, D.~Schlatter, I.~Schmitt and P.~M.~Zerwas,
  %``Scaling Violations in Inclusive e+ e- Annihilation Spectra,''
  Phys.\ Rev.\ D {\bf 27}, 105 (1983).
 % doi:10.1103/PhysRevD.27.105
  %%CITATION = doi:10.1103/PhysRevD.27.105;%%
  %2062 citations counted in INSPIRE as of 06 Jun 2017

\end{thebibliography}
\end{document}